\documentclass[11pt]{elsarticle}

\usepackage{lineno,hyperref}

\usepackage{float}
\usepackage{graphicx}
\usepackage{subfig}
\usepackage{mathrsfs}
\usepackage{graphics}
\usepackage{float}
\usepackage{amsmath}
\usepackage{amsfonts}
\usepackage{mathrsfs}
\usepackage{amsmath}

\usepackage{optidef}

\usepackage{optidef}
\usepackage{graphicx}
\usepackage{float}
\usepackage{graphicx}
\usepackage{subfig}
\usepackage{mathrsfs}
\usepackage{float}
\usepackage{amsmath}
\usepackage{amsfonts}
\usepackage{booktabs}
\usepackage{natbib}
\usepackage{bbm}
\usepackage{algorithm}
\usepackage[noend]{algpseudocode}
\usepackage{multirow}
\algdef{SE}[DOWHILE]{Do}{doWhile}{\algorithmicdo}[1]{\algorithmicwhile\ #1}

\usepackage[letterpaper, portrait, margin=1in]{geometry}

\usepackage{booktabs}
\usepackage{tabularx}

\usepackage{xcolor}
\hypersetup{
    colorlinks,
    linkcolor={red!50!red},
    citecolor={blue!50!blue},
    urlcolor={blue!80!blue}
}

\usepackage{optidef}

\usepackage{amsthm}

\theoremstyle{definition}

\usepackage{newtxtext} 

\DeclareMathOperator*{\argmax}{arg\,max}

\makeatletter
\newcommand*{\centerfloat}{%
  \parindent \z@
  \leftskip \z@ \@plus 1fil \@minus \marginparwidth
  \rightskip \leftskip
  \parfillskip \z@skip}
\makeatother

\usepackage{amssymb,amsmath,caption}
\usepackage[hang,flushmargin]{footmisc}
\newcommand{\algorithmfootnote}[2][\footnotesize]{
  \let\old@algocf@finish\@algocf@finish
  \def\@algocf@finish{\old@algocf@finish
    \leavevmode\rlap{\begin{minipage}{\linewidth}
    #1#2
    \end{minipage}}%
  }%
}

\usepackage{setspace}
\usepackage{natbib}
\usepackage{algorithm}
\usepackage[noend]{algpseudocode}
\algdef{SE}[DOWHILE]{Do}{doWhile}{\algorithmicdo}[1]{\algorithmicwhile\ #1}

\usepackage{appendix}

\usepackage{array}
\usepackage{threeparttable}
\usepackage{multirow}
\usepackage{rotating}
\usepackage{makecell}
\usepackage{longtable}

\journal{}

\biboptions{authoryear}

\begin{document}
\begin{frontmatter}

\title{Passenger Path Choice Estimation Using Smart Card Data: A Latent Class Approach with Panel Effects Across Days}


\author[label1]{Baichuan Mo}
\author[label5]{ZhenLiang Ma\corref{mycorrespondingauthor}}
\author[label3]{Haris N. Koutsopoulos}
\author[label4]{Jinhua Zhao}
\address[label1]{Department of Civil and Environmental Engineering, Massachusetts Institute of Technology, Cambridge, MA 02139}
\address[label5]{Department of Civil and Architectural Engineering, KTH Royal Institute of Technology, Stockholm 10044, Sweden}
\address[label3]{Department of Civil and Environmental Engineering, Northeastern University, Boston, MA 02115}
\address[label4]{Department of Urban Studies and Planning, Massachusetts Institute of Technology, Cambridge, MA 20139}

\cortext[mycorrespondingauthor]{Corresponding author}

\begin{abstract}
Understanding passengers’ path choice behavior in urban rail systems is a prerequisite for effective operations and planning. The area witnesses active developments in two broad but separate fields, including behaviour modeling using ‘small’ survey data in transport and mobility pattern using ‘big’ data in computer science. This paper attempts bridging the gap by proposing a probabilistic approach to infer passengers' path choice behavior in urban rail systems using a large-scale smart card data. The model uses latent classes and panel effects to capture passengers' implicit behavior heterogeneity and longitudinal correlations, key research gaps in big data driven behavior studies. We formulate the probability of each individual's arrival time at a destination based on their path choice behavior, and estimate corresponding path choice model parameters as a maximum likelihood estimation problem. The original likelihood function is intractable due to the exponential computation complexity. We derive a tractable likelihood function and propose a numerical integral approach to efficiently estimate the model. Also, we propose a method to calculate the t-statistic of the estimated choice parameters based on the numerically estimated Hessian matrix and Cramer–Rao bound (the lower bound on the coefficient variance). Case studies using synthetic data validate the model performance and its robustness against parameter initialization and input errors, and highlight the importance of incorporating crowding impact in path choice estimation. Applications using actual data from the Mass Transit Railway, Hong Kong reveal two latent groups of passengers: time-sensitive (TS) and comfort-aware (CA). TS passengers are those who are more likely to choose paths with short travel times. Most of them are regular commuters with high travel frequency and less schedule flexibility. CA passengers care more about the travel comfort experience and choose paths with less walking and waiting times. The proposed approach is data-driven and general to accommodate other discrete choice structures. It provides the same outputs as traditional choice modeling and facilities a deep understanding of passengers choice behaviors in both a cost-effective and timely way, based on which more informed planning and management strategies could be designed, evaluated, and monitored.  
\end{abstract}


\end{frontmatter}


\section{Introduction}\label{intro}

Increases in ridership are outpacing capacity in many large urban rail transit systems, including Hong Kong’s Mass Transit Railway (MTR), the London Underground, and the New York subway system \citep{zhu2017probabilistic,zhu2017inferring}. Crowding at stations and on trains is a concern due to its impact on safety, service quality, and operating efficiency. Various studies have measured passengers’ willingness to pay for less crowded conditions \citep{li2011crowding} and suggested incorporating the crowding disutility in investment appraisals \citep{haywood2015distribution}. Given the interest in dealing with crowding-related problems, understanding passengers’ route choice behavior under crowding situations is important for both operations management and planning practices. 
However, estimating path choice fractions or individual choice behavior is not trivial. As passenger's path choices in an urban rail system are not directly observed, most of the previous studies rely on revealed and stated preference survey data \citep{raveau2011topological, raveau2014behavioural, jin2017metro, zhang2017constrained}. Surveys are a powerful tool to facilitate behavior analysis. However, they are constrained by high costs, reporting accuracy, and survey coverage. 

Automated Fare Collection (AFC) and Automatic Vehicle Location (AVL) data provide opportunities for analysis in areas such as travel behavior, demand modeling, transit operations planning, etc. \citep{pelletier2011smart, bagchi2005potential, koutsopoulos2019transit}. In addition to aggregate trends of when and where passengers travel, AFC data provides detailed information on the travel patterns of individuals and/or specific groups \citep{goulet2016inferring, briand2017analyzing}. Table \ref{tab_liter} summarizes the existing route choice studies using AFC or/and AVL data in metro systems. Several studies have used AFC data to estimate passengers’ path choice probabilities \citep{sun2012rail, zhao2016estimation, sun2016schedule, zhou2015estimation, zhu2021passenger, mo2022assignment}. They provide useful insights on the aggregate choice behavior (i.e., path fractions) under existing conditions. For example, \citet{zhu2021passenger} develops a data-driven approach for the inference of passenger itineraries in urban heavy rail systems, where the path fractions can be estimated using AFC and AVL data. However, these results cannot be used for operations planning applications without modeling the individual path choice behavior, such as timetable design, network expansion, operating strategies and policy interventions, etc. This is because the new timetable or network expansion may change the service attributes, causing changes in an individual's choice behavior. Inference of path fractions cannot capture the impact of these changes.

This study focuses on the estimation of path choice models as a function of attributes of alternative paths using AFC and AVL data. Relevant to this context, \citet{sun2015integrated} developed an integrated Bayesian approach to infer network attributes and passenger route choice behavior using AFC data from the Singapore Mass Rapid Transit system. \citet{zhang2018estimating} developed a data fusion model to estimate individual path choices by combining revealed preference (RP) survey data and AFC data and modeled the risk attitudes of passengers. However, both studies imposed a strong assumption on link travel times (independent normal distribution) ignoring the fact that under congested conditions passengers may experience left behind at major stations due to capacity constraints. During peak periods, a (usually) shorter travel time route may have passengers who are left behind. But the models above cannot distinguish whether the longer travel time is due to choosing a longer route or being left behind multiple times on a shorter route \citep{zhu2017inferring,zhu2021passenger, mo2020capacity}. 

To incorporate the left behind phenomenon, \citet{zhu2017probabilistic} proposed a passenger-to-train assignment model (PTAM) by decomposing the journey time into access, waiting, in-vehicle, egress walking times, and considering the dynamics of being left behind at origin stations explicitly. The model was applied to estimate the left behind at key stations for non-transfer trips with capacity constraints and validated using both synthetic and actual data. \citet{horcher2017crowding} extended the PTAM to the case with transfers and presented a discrete choice model (DCM) to estimate the user cost of crowding in urban rail systems. However, the model identified the ``actual'' path used by passengers based on predetermined probability thresholds, which may eventually impact the estimation quality of the choice model. 

\begin{table*}[t]
\caption{Summary of literature on urban rail path choice inference and modeling using AFC data}
\label{tab_liter}
\resizebox{\linewidth}{!}{%
\begin{tabular}{@{}cccccccccc@{}}
\toprule
Reference                                    & \multicolumn{2}{c}{Data}                              & \multicolumn{2}{c}{Behavior}                          & \multicolumn{3}{c}{Characteristics}                                               & \multicolumn{2}{c}{Method}                            \\ \cmidrule(l){2-10} 
                                             & AFC                       & AVL                       & Aggregate                 & Individual                & Crowding                  & Heterogeneity            & Panel effect              & Optimization              & Probabilistic              \\ \midrule
\citet{sun2012rail}         & \checkmark & \checkmark & \checkmark &                           & \checkmark &                           &                           &                           & \checkmark \\
\citet{zhao2016estimation}  & \checkmark & \checkmark & \checkmark &                           & \checkmark &                           &                           &                           & \checkmark \\
\citet{sun2016schedule}     & \checkmark & \checkmark & \checkmark &                           & \checkmark &                           &                           &                           & \checkmark \\
\citet{zhou2015estimation}  & \checkmark & \checkmark & \checkmark &                           & \checkmark &                           &                           &                           & \checkmark \\
\citet{zhu2021passenger}    & \checkmark & \checkmark & \checkmark &                           & \checkmark &                           &                           &                           & \checkmark \\
\citet{mo2022assignment}    & \checkmark & \checkmark & \checkmark &                           & \checkmark &                           &                           & \checkmark &                           \\
\citet{sun2015integrated}   & \checkmark &                           &                           & \checkmark &                           &                           &                           &                           & \checkmark                          \\
\citet{zhang2018estimating} & \checkmark &                           &                           & \checkmark &                           &  &                           &                           & \checkmark                          \\
\citet{horcher2017crowding} & \checkmark & \checkmark &                           & \checkmark & \checkmark &                           &                           &                           & \checkmark \\ \midrule
\textbf{This study}         & \checkmark & \checkmark &                           & \checkmark & \checkmark & \checkmark & \checkmark &                           & \checkmark \\ \bottomrule
\end{tabular}
}
\end{table*}

These data-driven behavioral studies provide a good attempt to bridge the gap between 'small' (survey) and 'big' (AFC) data studies in different domains but answering the same question regarding passengers' path choices under crowding \citep{CHEN2016285}. However, existing AFC/AVL data driven path choice estimation studies are basically designed towards calibrating parameters of standard discrete choice models in DCM studies. They lack of systematic consideration of common and unique characteristics of the behavior choice problem itself and model them correspondingly in the context of big data, comparing with the survey data based DCM studies, including:
\begin{itemize}
    \item \textbf{Choice Heterogeneity}. It is commonly known that passengers may have different perceptions of service performance (i.e., travel time, waiting time) and show different choice strategies for travels. Ignoring the population heterogeneity may lead to estimation bias. The choice heterogeneity is usually captured by latent class or mixture models in the DCM literature \citep{hess2009taste, mo2021impacts}. However, to the best of the authors' knowledge, none of the previous AFC data-based studies have considered passenger heterogeneity in modeling individual path choice behavior.
    \item \textbf{Panel Effect (choice correlations across time)}. AFC data records passengers travels across days, and thus one unique challenge is about which days data should be used to estimate passengers' routine behavior. Considering individual travels on multiple days is important for robust choice behavior estimation, as passengers may occasionally deviate from their habitual behavior on some days. However, this brings challenges to model the temporal correlations of individuals' route choices (i.e., panel effect) across times and days, which is not considered or even discussed in previous studies.   
    \item \textbf{Choice Model Coefficient Significance (t-statistics)}. In DCM studies, the significance levels of model coefficients are important for deriving behavioral and policy insights. However, no AFC/AVL driven study is found on reporting the t-statistics of calibrated model coefficients, thus limiting its capability in facilitating comprehensive behavioral interpretation. 
\end{itemize}

To fill the research gaps, this paper develops a latent class approach with panel effects to estimate individual path choice behavior from AFC (tap-in and tap-out) and AVL data. The proposed framework explicitly captures capacity constraints (crowding), choice heterogeneity, and panel effects. We formulate the probability of each individual's arrival time at its destination station based on their path choice behavior, and estimate the corresponding path choice parameters as a maximum likelihood estimation (MLE) problem. The original likelihood function is intractable due to the exponentially large number of summations and the integration over a normally distributed variable. We derive a new conditional probability-based formulation to eliminate a large number of summations and use a numerical integration approach for the normal random variable, which leads to a tractable likelihood function and enables an efficient model estimation. Given the difficulty in deriving the analytical Hessian matrix, the t-values of estimated parameters are calculated based on the numerically estimated Hessian matrix and the Cramer–Rao bound. Case studies using synthetic data validate the model performance and highlight the importance of incorporating crowding impact in path choice estimation. Applications using actual data from the Mass Transit Railway (MTR), Hong Kong reveal two latent groups of passengers in the systems. The main contributions of this study are as follows: 
\begin{itemize}
    \item Introducing and model the passenger path choice problem using smart card data in closed public transport systems considering system crowding, choice heterogeneity and panel effects across times. 
    \item Formulating a MLE based latent-class path choice estimation problem with panel effects and deriving a tractable likelihood function for efficient model coefficients estimation. 
    \item Proposing a numerical method to calculate the t-statistic of estimated choice coefficients based on numerically estimated Hessian matrix and Cramer–Rao bound (lower bound of coefficient variance).
    \item Validating the model performance using both synthetic and real-world data in Hong Kong, and identifying latent groups of passengers with heterogeneous preference over travel times and comfortableness. 
\end{itemize}

The rest of the paper is organized as follows: Section \ref{sec_method} formulates the route choice problem and develops the MLE estimation method. Section \ref{sec_case} validates the proposed approach using synthetic and real-world data. The final section summarizes the main findings and discusses future directions.


\section{Methodology}\label{sec_method}
\subsection{Problem description}
Consider a closed AFC system where both tap-in and tap-out records of passengers over time are available, and train arrivals and departures at stations are available from the AVL system. Define a passenger $i$ with a series of observed AFC records $ v_i = \{ (o_{i,1},d_{i,1},t_{i,1}^{\text{in}},t_{i,1}^{\text{out}}), ..., (o_{i,N_i},d_{i,N_i},t_{i,N_i}^{\text{in}},t_{i,N_i}^{\text{out}})   \}$, where $o_{i,n},d_{i,n},t_{i,n}^{\text{in}},t_{i,n}^{\text{out}}$ represent the passenger's origin, destination, tap-in time, and tap-out time of the $n$-th trip, respectively. The set of all passengers is defined as $\mathcal{P}$ (i.e., $i \in \mathcal{P}$). 

To capture passengers' choice heterogeneity, we assume that there are $K$ latent groups in the population and passengers in the same group share the same choice preferences. Let $g_i$ be a random variable indicating the group that passenger $i$ belongs to. The probability that passenger $i$ belongs to a latent group $G_k$ is formulated as a multinomial logit model:
\begin{align}
\text{Pr}(g_i = G_k; \theta) = \frac{\exp(\theta_k \cdot x_i)}{\sum_{k'=1}^{K} \exp(\theta_{k'} \cdot x_i)} 
\label{eq_group}
\end{align}
where $x_i$ is the vector of the characteristics of passenger $i$, including variables (extracted from smart card data) such as travel frequency, card type, travel regularity, etc. $\mathcal{G} = \{G_k\;|\; k=1,2,...,K\}$ is a set of latent groups to be estimated ($K$ need to be pre-specified). $\theta = (\theta_k)_{k=1,...,K}$ is the parameter vector to be estimated, associated individual's characteristics.

According to the random utility maximization (RUM) assumption \citep{ben2018discrete}, the utility of passenger $i$ choosing path $m$ at the $n$-th trip, given that passenger $i$ is in group $G_k$, can be formulated as:
\begin{align}
U_{i,n,m}^k = \beta^k \cdot z_{n,m} + \alpha_{i}^k + \varepsilon_{i,n,m}^k
\label{eq_utility}
\end{align}
where $\beta^{k}$ are the unknown parameters to be estimated, associated with path attributes. $z_{n,m} := [y_{n,m},\;\log PS_m]$ and $y_{n,m}$ is the vector of path attributes, including variables such as in-vehicle time, out-of-vehicle time, left behind waiting time, etc. $PS_m$ is the ``path size'' factor, which is used to capture the correlation in error terms caused by path overlapping \citep{hoogendoorn2007modeling}. The formulation with the path size factor is known as the ``path-size logit model'' \citep{prato2009route}. $PS_m$ is defined as 
\begin{align}
PS_m & = \frac{1}{L_k}\sum_{a \in A_m} \frac{l_a}{\sum_{m'\in \mathcal{R}_{i,n}} \delta_{a, m'}}  \quad \forall \;  m \in \mathcal{R}_{i,n},  n = 1,...,N_i, i \in \mathcal{P}
\end{align}
where $A_m$ is the set of all links of path $m$. $\delta_{a, m'} = 1$ if link $a$ is in path $m'$, otherwise $\delta_{a, m'} = 0$. $L_k$ is the length of path $m$ and $l_a$ is the length of link $a$. $\mathcal{R}_{i,n}$ is the set of all available paths for passenger $i$'s $n$-th trip.

To capture the individual's behavior correlation over time (i.e., panel effect), the utility function (Eq. \ref{eq_utility}) also includes an individual specific unobserved factor $\alpha_{i}^k$ (a random variable). The panel effect is assumed to be persistent over time (i.e., no subscript $n$)  \citep{ben2018discrete}. $\alpha_{i}^k$ is assumed to be independent and identically distributed (i.i.d.) for all passengers in group $k$ and follows a normal distribution $\mathcal{N}(0, (\sigma^k)^2)$ (the zero-mean is due to the fact that the mean value can be estimated as a part of the alternative specific constant), where $\sigma^k$ is the standard deviation to be estimated. Given $\alpha_{i}^k$, the unobserved error term $\varepsilon_{i,n,m}^k$ is assumed to be i.i.d. Gumbel distributed across all $i$, $n$, and $m$. 

Let $ \pi^{k}_{i,n,m}[{\alpha_{i}^k}]$ be the probability of passenger $i$ choosing path $m$ at the $n$-th trip given that passenger $i$ is in group $G_k$. According to RUM theory,
\begin{align}
 \pi^{k}_{i,n,m}[{\alpha_{i}^k}] = \frac{\exp(\beta^k \cdot   z_{n,m} + \alpha_{i}^k )}{\sum_{m'\in \mathcal{R}_{i,n}} \exp(\beta^k \cdot   z_{n,m'} + \alpha_{i}^k)} 
\end{align}
Since there are a total of $N_i$ trip records for passenger $i$, we can formulate the series choice probability as \citep{arellano2001panel}:
\begin{align}
\text{Pr}(r_{i,1} = m_1,...,r_{i,N_i} = m_{N_i}) = \sum_{k=1}^{K}\text{Pr}(g_i = G_k) \cdot  \int_{\alpha_{i}^k} \left[\prod_{n=1}^{N_i}  \pi^{k}_{i,n,m_n}[{\alpha_{i}^k}] \right] \cdot f(\alpha_{i}^k)\text{ d} \alpha_{i}^k \nonumber \\
\quad \forall m_1\in \mathcal{R}_{i,1},...,m_{N_i}\in \mathcal{R}_{i,N_i}
\label{eq_P_all_choice}
\end{align}
where $r_{i,n}$ is a random variable indicating the path used by passenger $i$ in the $n$-th trip. And $f(\alpha_{i}^k)$ is the probability density function of $\alpha_{i}^k$.

The goal of this study is to develop an approach to simultaneously estimate $\beta = (\beta^k)_{k=1,...,K}$, $\theta$, $\sigma = (\sigma^k)_{k=1,...,K}$, which specify passenger's path choice behavior, choice heterogeneity, and panel effect. We formulate an MLE problem to estimate these parameters in the following sections.

\begin{figure}[htb]
\centering
\includegraphics[width=0.9 \linewidth]{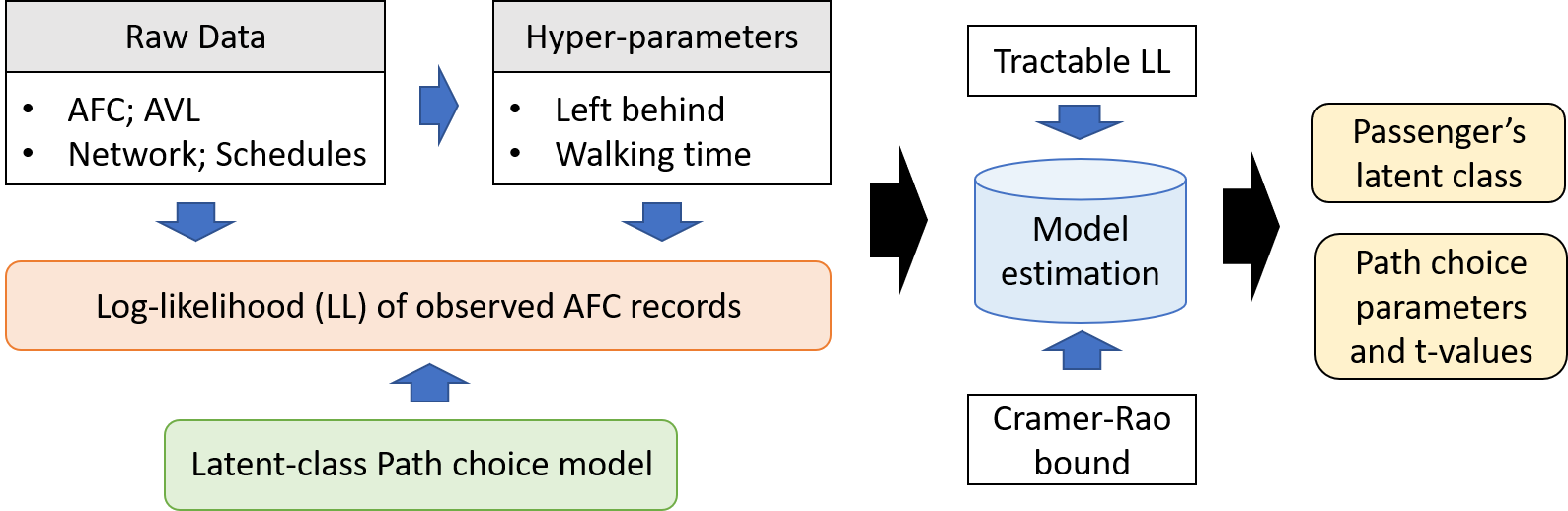}
\caption{Methodology framework}
\label{fig_frame}
\end{figure}

The structure of the methodology is presented in Figure \ref{fig_frame}.

The notation used across this paper is shown in Table \ref{tab:notation}.

\begin{longtable}{ p{.2\textwidth} | p{.8\textwidth} }
\caption{Notation summary}
\label{tab:notation}\\
\hline
{\textbf{Notation}} & {\textbf{Description}} \\
\toprule
\multicolumn{2}{l}{\emph{Model Parameters}}\\\midrule
$v_i$ & A series of AFC data records for passenger $i$ \\
$(o_{i,n},d_{i,n},t_{i,n}^{\text{in}},t_{i,n}^{\text{out}})$ & Passenger's origin, destination, tap-in time, and tap-out time of the $n$-th trip, respectively\\
$\mathcal{P}$ & The set of all passengers\\
$x_i$ & Vector of the characteristics of passenger $i$\\
$N_i$ & Total number of trips for passenger $i$\\
$G_k$ & The $k$-th latent group\\
$\mathcal{G}$ & The set of all latent groups\\
$K$ & The number of latent groups\\
$z_{n,m}$ & Vector of path attributes for path $m$ and trip $n$\\
$\alpha_{i}^k$ & A random variable to capture panel effect for individual $i$ in latent group $k$ \\
$U_{i,n,m}^k$ & The utility of passenger $i$ choosing path $m$ at the $n$-th trip, given that passenger $i$ is in group $G_k$\\
$PS_m$ & Path size factor for path $m$\\
$A_m$ & The set of all links of path $m$\\
$\mathcal{R}_{i,n}$ & The set of all available paths for passenger $i$'s $n$-th trip\\
$\tau$ & The time duration that each time index represents \\
$\mathcal{R}^{u,v}$ & The set of feasible paths for OD pair $(u,v)$\\
$ \pi^{k}_{i,n,m}[{\alpha_{i}^k}]$ & The probability of passenger $i$ choosing path $m$ at the $n$-th trip given that passenger $i$ is in group $G_k$\\
$r_{i,n}$ & A random variable indicating the path used by passenger $i$ in the $n$-th trip \\
$\mathcal{LL}(\theta, \beta, \sigma)$ & Log-likelihood function of all observations \\
$f(\alpha_{i}^k)$ & The probability density function of $\alpha_{i}^k $\\
$\Lambda_{i,n,m}^{j}$ & the set of all trains associated with the $j$-th segment of path $m$ for passenger $i$'s trip $n$\\ 
$J_{i,n,m}$ & Number of path segments for path $m$ of passenger $i$'s trip $n$\\
$\Omega_{i,n,m}$ & The set of possible itineraries for path $m$ in the $n$-th trip of passenger $i$ \\
$T_d(\cdot)$ & A function which returns the train's departure time at the boarding (resp. alighting) station of the corresponding segment \\
$T_a(\cdot)$ &A function which returns the train's arrival time at the boarding (resp. alighting) station of the corresponding segment \\
$f_m^{\text{Eg}}(\cdot)$ & Egress walking time probability density function (PDF) for path $m$ \\ 
$f_m^{\text{Ac}}(\cdot)$ & Access walking time PDF for path $m$ \\
$B_{i,n,m}^{I_1}$ & Maximum number of times that passenger $i$ is left behind to board Train $I_1$ in trip $n$ for path $m$ \\
$\eta_{i,n,m}^{j,k}$ & The probability of being left behind $k$ times at the boarding station of the $j$-th segment of path $m$ for passenger $i$'s trip $n$  \\   
$\mathcal{I}^{u,v,r}$ & The set of  legs for path $r$ of OD pair $(u,v)$\\ 
$E_{i,n,m}^{I_j,k}$ & The event that ``passenger $i$ in the $n$-th trip arrives at the boarding station of segment $j$ of path $m$ between the departure of Train $I_j - k$ and $I_j - k-1$ and is left behind $k$ times to board Train $I_j$''\\
$t_{i,n,m}^{j}$ & The transfer walking time from the alighting of train $I_{j-1}$ to the next platform for passenger $i$'s $n$-th trip using path $m$\\
$\hat{H}^{-1}_{k,k}$ & $k$-th diagonal element the inverse Hessian matrix for the log-likelihood function\\
\midrule
\multicolumn{2}{l}{\emph{Parameters to estimate}}\\
\midrule
$\beta^{k}$ & Parameters associated with path attributes in latent group $G_k$\\
$\sigma^k$ & The standard deviation of $\alpha_{i}^k$ \\
$\theta^k$ & Parameter vector associated individual's characteristics in latent group $G_k$\\
\bottomrule
\end{longtable}

\subsection{Model formulation}
Given the set of passenger $i$'s AFC data records $v_i$, the probability of observing $v_i$ can be expressed as
\begin{align}
\text{Pr}(v_i) &= \sum_{r_{i,1},...,r_{i,N_i}} \text{Pr}(v_i\;|\;r_{i,1},...,r_{i,N_i})\text{Pr}(r_{i,1} ,...,r_{i,N_i})
\nonumber \\ 
&= \sum_{r_{i,1},...,r_{i,N_i}} \left\{ \left[\prod_{n=1}^{N_i} \text{Pr}(o_{i,n},d_{i,n},t_{i,n}^{\text{in}},t_{i,n}^{\text{out}}  \;|\; r_{i,n})\right] \times  \text{Pr}(r_{i,1} ,...,r_{i,N_i}) \right\}
\label{eq_pr_v}
\end{align}
where the second equality follows from the Bayesian theorem.

As the origin and destination are known given path $m$, $\text{Pr}(o_{i,n},d_{i,n},t_{i,n}^{\text{in}},t_{i,n}^{\text{out}}  \;|\; r_{i,n} = m)$ is equivalent to the probability that passenger $i$ enters the origin at time $t_{i,n}^{\text{in}}$ and exits the destination at time $t_{i,n}^{\text{out}}$ (denoted as $\text{Pr}(t_{i,n}^{\text{in}},t_{i,n}^{\text{out}}\;|\; r_{i,n} = m)$). Notice that 
\begin{align}
&\text{Pr}(t_{i,n}^{\text{in}},t_{i,n}^{\text{out}}\;|\; r_{i,n}= m) = \text{Pr}(t_{i,n}^{\text{out}}\;|\;t_{i,n}^{\text{in}}, r_{i,n}= m) \cdot \text{Pr}(t_{i,n}^{\text{in}}\;|\; r_{i,n}) \propto \text{Pr}(t_{i,n}^{\text{out}}\;|\;t_{i,n}^{\text{in}}, r_{i,n}= m) 
\end{align}
The ``proportional to'' is due to the fact that we do not model the tap-in time choices. Therefore, the likelihood function becomes  
\begin{align}
&\mathcal{L}(\theta, \beta, \sigma) = \prod_{i \in \mathcal{P}} \text{Pr}(v_i) = \prod_{i \in \mathcal{P}}\left[\sum_{r_{i,1},...,r_{i,N_i}} \prod_{n=1}^{N_i}\left[ \text{Pr}(t_{i,n}^{\text{out}}\;|\;t_{i,n}^{\text{in}}, r_{i,n})\right] \cdot\text{Pr}(r_{i,1} ,...,r_{i,N_i})\right]
\end{align}

The only unknown part in the likelihood function (Equation 8) is $\text{Pr}(t_{i,n}^{\text{out}}\;|\;t_{i,n}^{\text{in}}, r_{i,n} = m)$, which is the probability that passenger $i$ taps out at his/her destination at time $t_{i,n}^{\text{out}}$ given that he/she uses path $m \in \mathcal{R}_{i,n}$ and taps in at time $t_{i,n}^{\text{in}}$. It can be derived by integrating over different itinerary scenarios, where each scenario is associated with a specific walking, boarding, and left behind possibility \citep{zhu2021passenger}. 

\begin{figure}[H]
\centering
\includegraphics[width=1 \linewidth]{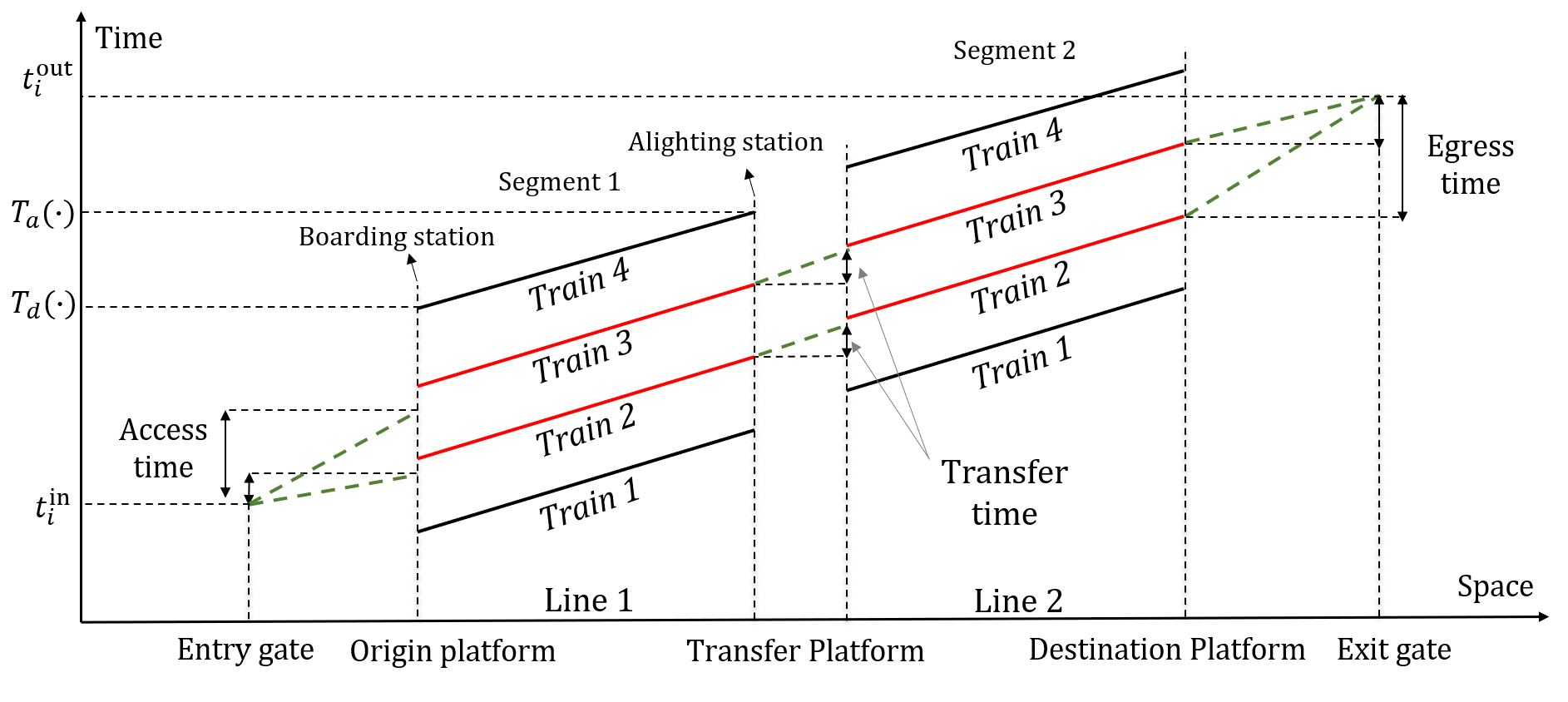}
\caption{Time-space diagram for a journey involving one transfer (adapted from \citet{zhu2017probabilistic}). The red lines indicate feasible itineraries}
\label{fig_ts_journey}
\end{figure}

To illustrate the derivation of $\text{Pr}(t_{i,n}^{\text{out}}\;|\;t_{i,n}^{\text{in}}, r_{i,n} = m)$, we consider an example journey involving one transfer. Figure \ref{fig_ts_journey} shows, in a time-space diagram, all possible movements of a passenger tapping in at the origin station on line 1 and tapping out at the destination station on line 2. The movement along a specific line is referred to as a ``path segment''. A path segment is characterized by the line, the boarding station, and the transfer/alighting station. Each path segment is associated with a set of trains with run IDs indicating the dispatching time sequence. For example, the first path segment in Figure \ref{fig_ts_journey} has Trains 1, 2, 3, and 4 numbered in chronological order. Let the set of all trains associated with the $j$-th segment of path $m$ for passenger $i$'s trip $n$ be $\Lambda_{i,n,m}^{j}$. For example, for the first path segment in Figure 1, we have $\Lambda_{i,n,m}^{j} = \{\emph{\text{Line 1 Train 1}},\; \emph{\text{Line 1 Train 2}},\; \emph{\text{Line 1 Train 3}},\; \emph{\text{Line 1 Train 4}}\}$. With slight abuse of notation, for a Train $I \in \Lambda_{i,n,m}^{j}$, Train $I + k$ represents the train in the same line with ID$+k$ ($k\in\mathbb{Z}$). For example, if Train $I$ is \emph{\text{Line 1 Train 1}}, then Train $I+1$ is \emph{\text{Line 1 Train 2}}.

After the passenger taps in, he/she walks directly to the platform at the origin station. The walking time from the entry gate to the origin station platform is referred to as the ``access walking time''. Depending on the available capacity (i.e., potentially left behind), this passenger may board Trains 2 or 3 on Line 1 for the first path segment. Note that Train 1 is not feasible because the passenger arrives on the platform after the departure of Train 1. After alighting at the transfer station, the passenger walks to the boarding platform for the next path segment on Line 2. The walking time from the alighting platform to the next boarding platform is referred to as the ``transfer time''. Similarly, depending on the available capacity, the passenger may board Trains 2 or 3 on Line 2 (Train 4 is not feasible because the passenger cannot exit at his/her current tap-out time if boarding Train 4). After alighting at the platform of the destination station, the passenger walks directly to the exit gate and taps out. The walking time from the alighting platform to the exit gate is referred to as the ``egress walking time''. 

Generally, let us consider passenger $i \in \mathcal{P}$ who uses path $m \in \mathcal{R}_{i,n}$ in his/her $n$-th trip. Let $J_{i,n,m}$ be the number of path segments for path $m$ of this trip. An itinerary $\mathcal{H} =\{I_1, I_2, ..., I_{J_{i,n,m}}\}$
is defined by ``a sequence of train IDs'' (each train ID is associated with a path segment) representing a possible movement of the passenger in the system, where $I_j \in \Lambda_{i,n,m}^{j}$ indicates Train $I_j$ for the $j$-th path segment. For example, in Figure \ref{fig_ts_journey}, a feasible itinerary is $\mathcal{H} = \{\emph{\text{Line 1 Train 2}},\; \emph{\text{Line 2 Train 3}}\}$, which indicates the itinerary that the passenger first boards Train 2 on Line 1 and then boards Train 3 on Line 2.

It is worth noting that for a specific passenger $i$, there are a limited number of feasible itineraries given his/her tap-in and tap-out time and the feasibility of transfer times. For example, in Figure \ref{fig_ts_journey}, any itineraries with trains in Line 1 departing before \emph{\text{Line 1 Train 2}} are not feasible because passengers cannot board those trains given their tap-in times. Let $\Omega_{i,n,m}$ be the set of possible itineraries for path $m$ in the $n$-th trip of passenger $i$. We have
\begin{align}
\Omega_{i,n,m} = \{\{I_1, ..., I_{J_{i,n,m}}\}, \;  \forall \; I_j \in \Lambda_{i,n,m}^{j},\;|\;  T_d(I_1) \geq t_{i,n}^{\text{in}}, 
T_a(I_{J_{i,n,m}}) \leq t_{i,n}^{\text{out}}, T_d(I_j) \geq T_a(I_{j+1}),\; \nonumber \\ \forall \; j = 1,..., J_{i,n,m} \}
\label{eq_feasible_I}
\end{align}
where $T_d(\cdot)$ (resp. $T_a(\cdot)$) is a function which returns the train's departure (resp. arrival) time at the boarding (resp. alighting) station of the corresponding segment. This information is available from the AVL data. Eq. \ref{eq_feasible_I} means that a feasible itinerary needs to satisfy that 1) Train $I_1$ departs after $t_{i,n}^{\text{in}}$ so that the passenger is able to board it (i.e., $T_d(I_1) \geq t_{i,n}^{\text{in}}$, assuming the minimum access walking time is zero). 2) The last train (i.e., Train $I_{J_{i,n,m}}$) arrives earlier than $t_{i,n}^{\text{out}}$ so that the passenger is able to tap-out at $t_{i,n}^{\text{out}}$ (i.e., $T_a(I_{J_{i,n,m}}) \leq t_{i,n}^{\text{out}}$, assuming the minimum egress walking time is zero). 3) Train $I_{j+1}$ departs later than the arrival of the train $I_{j}$ so that the passenger can successfully transfer (i.e., $T_d(I_j) \geq T_a(I_{j+1})$, assuming the minimum transfer time is zero).

Given the feasible itinerary set $ \Omega_{i,n,m}$, $\text{Pr}(t_{i,n}^{\text{out}}\;|\;t_{i,n}^{\text{in}}, r_{i,n} = m)$ can be rewritten as:
\begin{align}
&\text{Pr}(t_{i,n}^{\text{out}} \;|\;t_{i,n}^{\text{in}}, r_{i,n} = m ) =\sum_{\mathcal{H} \in \Omega_{i,n,m}} \text{Pr}(t_{i,n}^{\text{out}} \;|\;\mathcal{H}, t_{i,n}^{\text{in}}, r_{i,n} = m) \cdot \text{Pr}(\mathcal{H}\;|\;t_{i,n}^{\text{in}}, r_{i,n} = m).
\end{align}

We first consider the derivation of $\text{Pr}(t_{i,n}^{\text{out}} \;|\;\mathcal{H}, t_{i,n}^{\text{in}}, r_{i,n} = m)$, the probability of tap out at time $t_{i,n}^{\text{out}}$ given itinerary $\mathcal{H}$, path $m$ and tap-in time. Since the itinerary includes the information of the last train's arrival time $T_a(I_{J_{i,n,m}})$, this probability is equivalent to the probability that the egress walking time is equal to $t_{i,n}^{\text{out}} - T_a(I_{J_{i,n,m}})$. Let the egress walking time probability density function (PDF) for path $m$ be $f_m^{\text{Eg}}(\cdot)$. Then, $\text{Pr}(t_{i,n}^{\text{out}} \;|\;\mathcal{H}, t_{i,n}^{\text{in}}, r_{i,n} = m)$ can be expressed as
\begin{align}
\text{Pr}(t_{i,n}^{\text{out}} \;|\;\mathcal{H}, t_{i,n}^{\text{in}}, r_{i,n} = m) = f_m^{\text{Eg}}(t_{i,n}^{\text{out}} - T_a(I_{J_{i,n,m}})).
\label{eq_t_out}
\end{align}
Note that Equation \ref{eq_t_out} uses the density to represent the probability, which does not affect the parameter estimations in the MLE.

Now let us consider the derivation of $\text{Pr}(\mathcal{H}\;|\;t_{i,n}^{\text{in}}, r_{i,n} = m)$, the probability of choosing itinerary $\mathcal{H}$ given path $m$ and the tap-in time. Since the boarded train on segment $j$ only depends on the boarded train on segment $j-1$, but not $j-k$ for all $k>1$, this probability can be extended using the Markov property:
\begin{align}
\text{Pr}(\mathcal{H}\;|\;t_{i,n}^{\text{in}}, r_{i,n} = m) &= \text{Pr}({I_1,I_2,...,I_{J_{i,n,m}}}\;|\;t_{i,n}^{\text{in}}, r_{i,n} = m) \nonumber\\
&=\text{Pr}(I_1\;|\;t_{i,n}^{\text{in}}, r_{i,n} = m) \cdot \prod_{j=2} ^{J_{i,n,m}}\text{Pr}(I_j \;|\;I_{j-1},t_{i,n}^{\text{in}}, r_{i,n} = m)
\label{eq_Pr_I}
\end{align}
In the following contents, we elaborate the derivation of $\text{Pr}(I_1\;|\;t_{i,n}^{\text{in}}, r_{i,n} = m)$ and $\text{Pr}(I_j \;|\;I_{j-1},t_{i,n}^{\text{in}}, r_{i,n} = m)$, respectively.

Note that $\text{Pr}(I_1\;|\;t_{i,n}^{\text{in}}, r_{i,n} = m)$ is the probability of boarding Train $I_1$ on the first segment of path $m$. There are two different scenarios for this event to happen: 1) \textbf{[No left behind]} the passenger arrives at the platform between the departure time of Trains $I_1$ and $I_1 - 1$ and boards Trains $I_1$ without left behind. 2) \textbf{[With left behind]} the passenger arrives at the platform between the departures of Trains $I_1 - k$ and $I_1 - k-1$ and is able to board Train $I_1$ after being left behind $k$ times. Given the feasible itinerary set $\Omega_{i,n,m}$, there is a maximum number of times the passenger is left behind to board Train $I_1$. Denote the upper bound of $k$ as $B_{i,n,m}^{I_1}$, where $B_{i,n,m}^{I_1} = \argmax_k \{k\in \mathbb{N}\;| \; \exists \; \mathcal{H}' \in \Omega_{i,n,m} \text{ s.t. } I'_1 = I_1-k, I'_1\in \mathcal{H}'\}$, $\mathbb{N}$ is the set of natural numbers (including zero). And Train $I_1 - B_{i,n,m}^{I_1}$ represents the earliest train that passenger $i$ can board at the first segment. 

Let $t_{i,n,m}^{j}$ be the walking time from the alighting platform of segment $j-1$ to the boarding platform of segment $j$ in path $m$ for passenger $i$, and $t_{i,n,m}^{0}$ is the access walking time. Then, $t_{i,n}^{\text{in}} + t_{i,n,m}^{0}$ is the passenger arrival time at the platform of his/her origin station. Hence, the probability of arriving at the platform between the departure of Train $I_1 - k$ and $I_1 - k-1$ can be formulated as 
\begin{align}
\text{Pr}(T_d(I_1-k-1) \leq t_{i,n}^{\text{in}} + t_{i,n,m}^{0} \leq T_d(I_1-k)\;|\;t_{i,n}^{\text{in}}, r_{i,n} = m)  = \int_{T_d(I_1-k-1) -  t_{i,n}^{\text{in}}}^{T_d(I_1-k) - t_{i,n}^\text{in}}  f_m^{\text{Ac}}(t) dt := \rho_{i,n,m}^{I_1,k} \nonumber\\ \forall k=0,1,..,B_{i,n,m}^{I_1}
\label{eq_rho}
\end{align}
where $f_m^{\text{Ac}}(\cdot)$ is the access walking time PDF for path $m$. Eq. \ref{eq_rho} can be pre-calculated once $f_m^{\text{Ac}}(\cdot)$ is given because it is a definite integral. 

Let $\eta_{i,n,m}^{j,k}$ be the probability of being left behind $k$ times at the boarding station of the $j$-th segment of path $m$ for passenger $i$'s trip $n$. Let $E_{i,n,m}^{I_j,k}$ be the event that ``passenger $i$ in the $n$-th trip arrives at the boarding station of segment $j$ of path $m$ between the departure of Train $I_j - k$ and $I_j - k-1$ and is left behind $k$ times to board Train $I_j$''. We have:
\begin{align}
\text{Pr}({E_{i,n,m}^{I_1,k}|\;t_{i,n}^{\text{in}}, r_{i,n} = m}) &= \rho_{i,n,m}^{I_1,k} \cdot \eta_{i,n,m}^{1,k} \quad \forall k=0,1,..,B_{i,n,m}^{I_1}.
\end{align}
Then, $\text{Pr}(I_1\;|\;t_{i,n}^{\text{in}}, r_{i,n} = m)$ can be rewritten as 
\begin{align}
\text{Pr}(I_1\;|\;t_{i,n}^{\text{in}}, r_{i,n} = m) &= \sum_{k=0}^{B_{i,n,m}^{I_1}}  \text{Pr}({E_{i,n,m}^{I_1,k}|\;t_{i,n}^{\text{in}}, r_{i,n} = m}) = \sum_{k=0}^{B_{i,n,m}^{I_1}}   \rho_{i,n,m}^{I_1,k} \cdot \eta_{i,n,m}^{1,k}.
\end{align}
This finishes the derivation of $\text{Pr}(I_1\;|\;t_{i,n}^{\text{in}}, r_{i,n} = m)$. $\eta_{i,n,m}^{1,k}$ can be estimated from AFC and AVL data using a Gaussian Mixture model \citep{ma2019estimation}, which will be described in \ref{sec_LB_est}.

Now, we derive $\text{Pr}(I_j \;|\;I_{j-1},t_{i,n}^{\text{in}}, r_{i,n} = m)$ in Eq. \ref{eq_Pr_I}, the probability of boarding train $I_j$ given that the passenger has boarded train $I_{j-1}$ on the $(j-1)$-th segment of path $m$. It is derived in a similar way as $\text{Pr}(I_1\;|\;t_{i,n}^{\text{in}}, r_{i,n} = m)$. Passenger $i$ may arrive at the boarding station of segment $j$ between the departure times of Trains $I_j - k$ and $I_j - k-1$ and be left behind $k$ times to board Train $I_j$ (note that $k=0$ means no left behind). The probability of arriving at the platform between the departures of train $I_j - k$ and $I_j -k -1$ given he/she alights at $T_a(I_{j-1})$ is formulated as
\begin{align}
&\text{Pr}(T_d(I_j-k-1) \leq T_a(I_{j-1}) + t_{i,n,m}^{j} \leq T_d(I_j-k)\;|\;I_{j-1}, t_{i,n}^{\text{in}}, r_{i,n} = m) \nonumber \\
& = \int_{T_d(I_j-k-1) -   T_a(I_{j-1})}^{T_d(I_j-k) -  T_a(I_{j-1})}  f_{m,j}^{\text{Tr}}(t) dt = \Tilde{\rho}_{i,n,m}^{I_j,k} \quad \quad \quad \forall k=0,1,..,B_{i,n,m}^{I_j}.
\end{align}
where $B_{i,n,m}^{I_j}$ is the maximum possible left behind times when boarding train $I_j$ given the feasible itinerary constraint, defined as $\argmax_k \{k\in \mathbb{N}\;| \; \exists  \mathcal{H}' \in \Omega_{i,n,m} \text{ s.t. } I'_j = I_j-k, I'_j  \in  \mathcal{H}'\}$. $t_{i,n,m}^{j}$ is the transfer walking time from the alighting of train $I_{j-1}$ to the next platform.  $f_{m,j}^{\text{Tr}}(\cdot)$ is the PDF of $t_{i,n,m}^{j}$.  $\Tilde{\rho}_{i,n,m}^{I_j,k}$ is defined for the simplicity of expression. Given the definition of $E_{i,n,m}^{I_j,k}$, we have
 \begin{align}
& \text{Pr}({E_{i,n,m}^{I_j,k}|\;I_{j-1}, t_{i,n}^{\text{in}}, r_{i,n} = m}) =\Tilde{\rho}_{i,n,m}^{I_j,k} \cdot \eta_{i,n,m}^{j,k} \quad \forall k=0,1,..,B_{i,n,m}^{I_j}.
\end{align}
Then, $\text{Pr}(I_j \;|\;I_{j-1},t_{i,n}^{\text{in}}, r_{i,n} = m)$ can be rewritten as
\begin{align}
&\text{Pr}(I_j\;|\;I_{j-1}, t_{i,n}^{\text{in}}, r_{i,n} = m) = \sum_{k=0}^{B_{i,n,m}^{I_j}}  \text{Pr}({E_{i,n,m}^{I_j,k}|\;I_{j-1}, t_{i,n}^{\text{in}}, r_{i,n} = m}) =\sum_{k=0}^{B_{i,n,m}^{I_j}}    \Tilde{\rho}_{i,n,m}^{I_j,k} \cdot \eta_{i,n,m}^{j,k}.
\end{align}

With all parts of $\mathcal{L}(\theta, \beta, \sigma)$ in Equation 8 derived, there are still two remaining challenges for the MLE problem. First, the calculation of $\text{Pr}(t_{i,n}^{\text{out}}\;|\;t_{i,n}^{\text{in}}, r_{i,n})$ requires the inputs of left behind probability $\eta_{i,n,m}^{j,k}$ and three PDF functions $f_m^{\text{Ac}}(\cdot)$, $f_m^{\text{Eg}}(\cdot)$, and $f_{m,j}^{\text{Tr}}(\cdot)$. The PDF functions can be obtained from field-experiments. But obtaining the left behind probability is not trivial. In this study, we used the model proposed by \citet{ma2019estimation} to estimate $\eta_{i,n,m}^{j,k}$ from AFC and AVL data. Details can be found in \ref{sec_LB_est}. The second challenge is that, the calculation of $\text{Pr}(v_i)$ (Eq. \ref{eq_pr_v}) has an exponentially large number of summation over different paths, and it requires the integral of a normally distributed random variable, which makes it numerically hard to solve. In the following section, we derive a new conditional probability-based formulation to eliminate the large number of summations, and 
use a numerical integration approach for the normal random variable, which leads to a tractable likelihood function and enables an efficient model estimation. 

\subsection{Tractable log-likelihood function}
To eliminate the exponentially large number of summation over different paths in Eq. \ref{eq_pr_v}, we observe that given $\alpha_i^k$ and $g_i$, passenger $i$'s route choice for each trip becomes independent. Mathematically,
\begin{align}
&\text{Pr}(v_i \;|\; \alpha_i^k, g_i = G_k) =  \prod_{n=1}^{N_i}  \text{Pr}(t_{i,n}^{\text{out}}, t_{i,n}^{\text{in}}\;|\; \alpha_i^k, g_i = G_k) = \prod_{n=1}^{N_i} \sum_{m_n \in \mathcal{R}_{i,n}}\text{Pr}(t_{i,n}^{\text{out}} \;|\; t_{i,n}^{\text{in}}, r_{i,n} = m_n) \cdot \pi^{k}_{i,n,m_n}[{\alpha_{i}^k}] 
\label{eq_new_v}
\end{align}
Note that Eq.\ref{eq_new_v} only has a total of $N_i\times|\mathcal{R}_{i,n}|$ summation terms, while this number in Eq.\ref{eq_pr_v} is $|\mathcal{R}_{i,n}|^{N_i}$. Based on Eq. \ref{eq_new_v}, $\text{Pr}(v_i)$ can be obtained by integrating and summing over $\alpha_i^k$ and $g_i$, respectively. Since $\alpha_i^k$ is a normal random variable, an approximated numerical integration approach is used to get a tractable formula. Note that there are a large class of quadrature rules for numerical integration \citep{davis2007methods}. In this paper, we use the  simplest midpoint rule for the interpolation as this is not the focus of this study. Let $\alpha^{\text{U}}$ and $\alpha^{\text{L}}$ be the upper and lower bounds for $\alpha_i^k$.
We divide $[\alpha^{\text{L}}$, $\alpha^{\text{U}}]$ into discrete intervals with equal length $\Delta$. Let $\mathcal{S}$ be the set of all middle points in each interval. Specifically, $\mathcal{S} = \{\alpha^{\text{L}} + \frac{k\cdot \Delta}{2} \mid \forall k = 1,3,5,...,|\mathcal{S}|, |\mathcal{S}| = \frac{2(\alpha^{\text{U}}-\alpha^{\text{L}})}{\Delta} - 1\}$.
Hence, $\text{Pr}(v_i)$ can be rewritten as
\begin{align}
\text{Pr}(v_i) \approx   \sum_{k=1}^{K}\text{Pr}(g_i = G_k)  \cdot  \sum_{\alpha_i^k \in \mathcal{S}}\text{Pr}(v_i \;|\; \alpha_i^k, g_i = G_k) \cdot f(\alpha_i^k) \cdot \Delta
\end{align}
$\Delta$ is the parameter determining the trade-off between approximation accuracy and computational efficiency, where a smaller $\Delta$ indicates a more fine-grained integration, but higher computational cost. 

Given the new formulation of $\text{Pr}(v_i)$, we can use $\mathcal{L}(\theta, \beta, \sigma) = \prod_{i \in \mathcal{P}} \text{Pr}(v_i) $ to evaluate the likelihood function with a tractable formulation.

\subsection{Model estimation}
The new log-likelihood function can be expressed as 
\begin{align}
\mathcal{LL}(\theta, \beta, \sigma) = \sum_{i \in \mathcal{P}} \log \text{Pr}(v_i)  =
\sum_{i \in \mathcal{P}} \log  \left[ \sum_{k=1}^{K}\sum_{\alpha_i^k \in \mathcal{S}} \text{Pr}(g_i = G_k)  \cdot  \text{Pr}(v_i \;|\; \alpha_i^k, g_i = G_k) \cdot f(\alpha_i^k) \cdot \Delta \right]
\end{align}
As $\mathcal{LL}(\theta, \beta, \sigma)$ is a combination of elementary functions, it is continuous and differentiable. Therefore, the MLE can be solved with any first- or second-order optimization method. In this study, the BFGS algorithm is used \citep{nocedal2006numerical}. BFGS is a quasi-Newton method. It uses only the first derivatives and has demonstrated good performance for many optimization problems. However, as the function includes the multiplication of several nonlinear terms, the convexity of this function is unknown. It is possible that the $\mathcal{LL}$ is not concave and the BFGS algorithm may converge to different local minimums under different initializations. Hence, we conduct a sensitivity analysis in Section \ref{sec_sentitivity} with respect to different initial values and show that the model estimation results are stable. Besides, the numerical results show that $\mathcal{LL}$ is concave within a reasonable range of path attribute values.

After obtaining the optimal parameters $\theta^*$, $\beta^*$, and $\sigma^*$, we calculate the t-values of the estimated parameters based on a numerically estimated Hessian matrix and the Cramer-Rao bound. Note that as $\mathcal{LL}$ is second-order differentiable, the analytical Hessian matrix can also be derived. The numerical Hessian matrix is used for simplification due to the complex function form. In this study, we adopt the formulation with fourth-order approximation under uniform grid spacing to calculate the second derivative \citep{fornberg1988generation}. The exact formulas are attached in \ref{sec_hessian} (other approximation formulas can also be used). With the second derivative formulas, we can calculate the numerical Hessian matrix of $\mathcal{LL}(\theta, \beta, \sigma)$ at point ($\theta^*, \beta^* ,\sigma^*$). Denote the Hessian matrix as $\hat{H}$. Note that, from the second-order optimality
conditions, $\hat{H}$ is negative semi-definite, which is the algebraic equivalent of the local concavity of the log-likelihood function \citep{bierlaire2020short}. 

Let $\Theta = (\theta, \beta, \sigma)$ be a vector of all parameters. Using the Cramer-Rao bound \citep{cramer2016mathematical, rao1992information}, the variance of an estimated parameter $\hat{\Theta}_k$ is 
\begin{align}
   {\text{Var}[\hat{\Theta}_k]} = -\hat{H}^{-1}_{k,k}
\end{align}
where $\hat{H}^{-1}$ is the inverse of the Hessian matrix and $\hat{H}^{-1}_{k,k}$ is its $k$-th diagonal element. Then, the corresponding t-value is calculated as:
\begin{align}
   {\text{t-value}[\hat{\Theta}_k]} = \frac{\hat{\Theta}_k}{\sqrt{\text{Var}[\hat{\Theta}_k]}}
\end{align}

\section{Case study}\label{sec_case}

\subsection{Model validation and sensitivity test}\label{sec_sentitivity}
It is difficult to collect passengers' actual path choices in reality. To validate the proposed approach, we use synthetic data generated by simulating passengers' route choices, train operations, and their interactions \citep{mo2022assignment,zhu2021passenger,mo2020estimating}. 

\begin{figure}[htb]
\centering
\includegraphics[width=0.6 \linewidth]{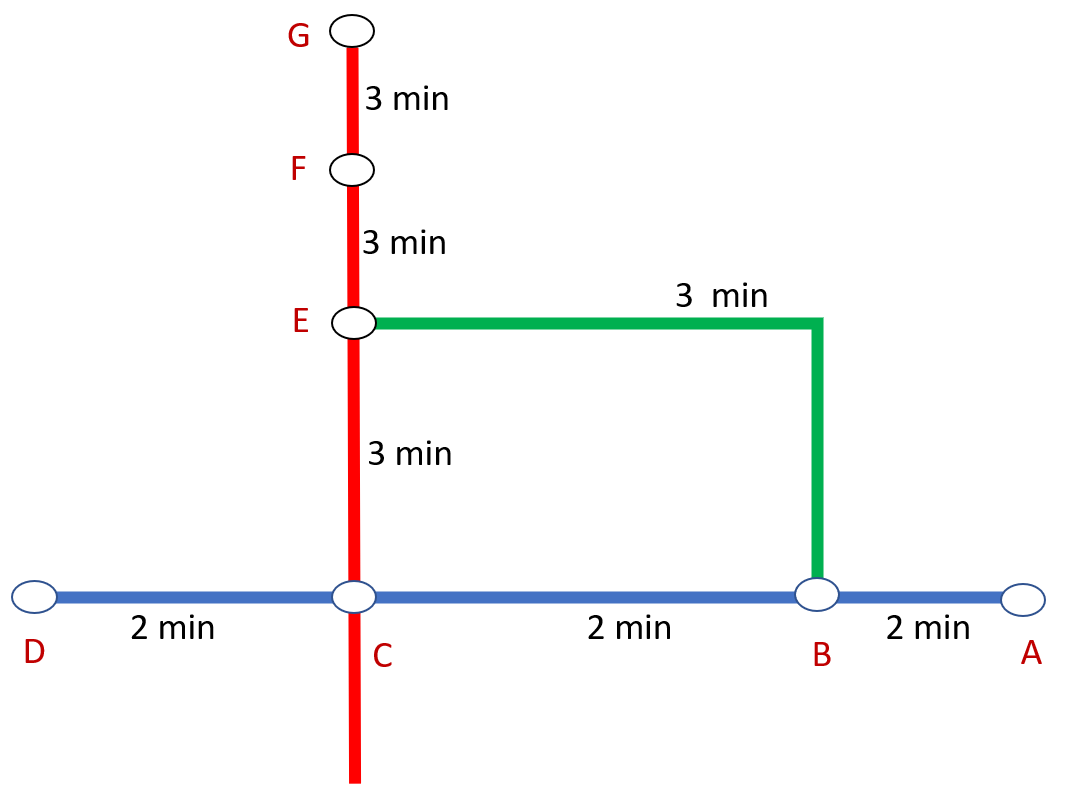}
\caption{Synthetic urban rail network}
\label{fig_syn_urban_net}
\end{figure}

Figure \ref{fig_syn_urban_net} shows the configuration of the synthetic urban rail network, where there are 7 stations (A$\sim$G) and 3 lines (red, green, and blue). The number on each link represents the in-vehicle travel time. This network is extracted from the MTR metro system in Hong Kong. It is also representative of typical metro network structures in terms of lines and transfers. The platform of station C of the red line in the up direction is assumed to be crowded with extensive left behind. All the other platforms are assumed to have no left behind.

To generate the synthetic data, we assume that passengers' path choice behavior is based on four path attributes: 1) in-vehicle time (i.e., the train run time of a path), 2) out-of-vehicle time (i.e., the sum of access, egress, transfer, and waiting time without left behind), 3) the number of transfers (i.e., the number of times transferring on the path), and 4) denied waiting time (i.e., the waiting time due to left behind at the crowded platforms). We also assume passenger's latent groups can be characterized by two sociodemographic variables $x^{(1)}$ and $x^{(2)}$, where $x^{(1)}$ is drawn from $\mathcal{U}[-4,4]$ and $x^{(2)}$ is drawn from $\mathcal{U}[-2,2]$. Suppose there are two latent groups for the synthetic passengers: ``time-sensitive'' (TS) and ``comfort-aware'' (CA). The TS passengers, when making path choices, tend to minimize their total travel time, meaning that the impact of in-vehicle time, out-of-vehicle time, and denied waiting time are similar to passenger's path choice utility. CA passengers prefer paths with less walking or waiting time though the in-vehicle time could be longer. That is, the out-of-vehicle time and denied waiting time have a higher impact on these passengers' path choice utilities than the in-vehicle time. 

Table \ref{tab_para} shows the parameters for the latent class path choice model for generating the synthetic data. These parameters are chosen based on the survey results in \citet{jin2017metro}. We set the in-vehicle time and path size factor parameters to be the same for TS and CA passengers according to the survey modeling results. In addition, we set the choice parameters to be $0.7\times$ ($1.5\times$) of the parameters from the survey results for the TS (CA) group. The parameters for the latent class model are set as 1.5 and 0.6 for $x^{(1)}$ and $x^{(2)}$, respectively. 

\begin{table}[H]
\caption{Path choice parameters for synthetic data generation}
\centering
\begin{tabular}{@{}cccc@{}}
\toprule
\multicolumn{1}{c}{Variables}           & \multicolumn{1}{c}{TS}    & \multicolumn{1}{c}{CA} & \multicolumn{1}{c}{\citet{jin2017metro}}    \\ \midrule
\multicolumn{4}{l}{\emph{Path choice parameters}}                                                                \\
\multicolumn{1}{c}{In-veh time}         & -0.2676 & -0.2676                 & -0.2676 \\
\multicolumn{1}{c}{Out-of-veh time}     & -0.2980                     & -0.6386                 & -0.4257                     \\
\multicolumn{1}{c}{Num of transfer  }     & -1.3068                     & -3.1737                 & -1.8669                     \\
\multicolumn{1}{c}{Denied waiting time}   & -0.3222                     & -0.7825                 & -0.4603                     \\
\multicolumn{1}{c}{$\log PS_m$ }               & 0.5815                      & 0.5815                  & 0.5815                      \\
\multicolumn{1}{c}{$\sigma^k$ }              & 1                           & 1                       & N/A                         \\ \midrule
\multicolumn{4}{l}{\emph{Latent class parameters}}                                                               \\
\multicolumn{1}{c}{$x^{(1)}$ }              & \multicolumn{3}{c}{1.5}                                                             \\
\multicolumn{1}{c}{$x^{(2)}$}              & \multicolumn{3}{c}{0.6}                                                             \\ \bottomrule
\end{tabular}
\label{tab_para}
\end{table}

Table \ref{tab_system_para} summarizes the parameters of the network, train operations, and passengers. The synthetic data is generated for 9 OD pairs (origin stations A, B, D, and destination stations E, F, G) by simulating the tap-out time given tap-in time. All the OD pairs have 2 paths. For example, the possible paths for OD pair (A, E) are A-B-E and A-B-C-E. Without loss of generality, we assume that there are 2,700 passengers, each of whom performs 3 trips (i.e., $N_i = 3$). Each trip is associated with a randomly selected OD pair. Algorithm \ref{alg_data_generation} describes the detailed synthetic data generation procedure.

\begin{table}[H]
\caption{System settings for synthetic data generation}
\centering
\resizebox{1\linewidth}{!}{
\begin{tabular}{@{}p{0.3\linewidth}p{0.7\linewidth}@{}}
\toprule
Entity           & Settings                                                                                                                                                                                                                                                                            \\ \midrule
Network          & Walk distance 30-50 meters (access, egress, transfer)                                                                                                                                                                                                                                     \\\midrule
Train operations & \begin{tabular}[c]{@{}l@{}}Headway 2+$\delta_H$ minutes.\\ $\delta_H$ is drawn uniformly from $[-10, 10]$ seconds\\   In-vehicle time$+\delta_V$ (see Figure \ref{fig_syn_urban_net})  \\ 
$\delta_V$ is drawn uniformly from $[-20, 20]$ seconds\end{tabular}                                                                                                                                  \\\midrule
Passengers       & \begin{tabular}[c]{@{}l@{}}Walk speed distribution follows a lognormal distribution\\ with mean of 1.2m/s and standard deviation of 0.5m/s.\\  Left behind probabilities at station C, red line,  up\\ direction are 20\% no left behind, 50\% left behind once\\  and 30\% left behind twice\end{tabular} \\ \bottomrule
\end{tabular}
}
\label{tab_system_para}
\end{table}

\begin{algorithm}
\small
\caption{Synthetic data generation}
\begin{algorithmic}[1]
\renewcommand{\algorithmicrequire}{\textbf{Input:}}
\renewcommand{\algorithmicensure}{\textbf{Output:}}
\Require Path choice parameters; Passenger set $\mathcal{P}$.
\Ensure  Synthetic AFC data (tap-in/tap-out stations/times)
\State Initialize the number of sample instances $N$. 
\For {$i \in \mathcal{P}$}
\State Sample  $x^{(1)}_i \sim  \mathcal{U}[-4,4]$, $x^{(2)}_i \sim  \mathcal{U}[-2,2]$.
\State Sample group $g_i \sim \text{Pr}(g_i ;\theta)$. Denote the group as $G_k$.
\State Sample $\alpha_i^k \sim \mathcal{N}(0, \theta^k)$.
\For {$n = 1$ to $N_i$}
\State Sample $t_{i,n}^{\text{in}} \sim \mathcal{U}[\text{7:00AM, 10:00AM}]$.
\State Randomly sample an OD pair for this trip. 
\State Calculate the path choice probability $ \pi^{k}_{i,n,m}[{\alpha_{i}^k}]$.
\State Sample a path $m \in \mathcal{R}_{i,n}$ based on $ \pi^{k}_{i,n,m}[{\alpha_{i}^k}]$.
\State Sample the actual travel information (i.e., train run time, headway, access, egress, transfer, and denied waiting times) for this trip based on path $m$. Obtain $t_{i,n}^{\text{out}}$.
\EndFor
\State Save $t_{i,n}^{\text{in}}$ and $t_{i,n}^{\text{out}}$ and the OD as a trip record.
\EndFor
\State Combine all trip records as the synthetic AFC data.
\end{algorithmic} 
\label{alg_data_generation}
\end{algorithm}

In total, 8,100 trips from the 2,700 passengers were generated. The synthetic AFC data is then used for model estimation. As we have the ``true'' value of choice parameters (Table \ref{tab_para}), we can validate the model's performance. The MLE is solved using the BFGS algorithm \citep{fletcher2013practical} in the Python \texttt{Scipy} package. $\alpha^{\text{L}} = -3$, $\alpha^{\text{U}} = 3$, and $\Delta = 1$ are used for numerical integration. 

Table \ref{tab_syn_results} shows the estimation results of the path choice parameters. The percentage values in the brackets quantify the relative errors compared to the ``true'' parameters.  Note that the actual walking speed distribution and left behind probabilities are used in the model estimation. And the sensitivity analysis on these inputs is shown in Section \ref{sec_sentitivity}. For comparison purposes, we also estimate a baseline model without latent classes. The latent class model can estimate the actual parameters with a mean percentage error of around 10\%. It outperforms the baseline model in estimation accuracy. The out-of-vehicle time parameter for the TS group has the maximum error (-33.2\%), which may be due to the fact that the out-of-vehicle time is highly correlated with the number of transfers, making the numerical estimation harder. Note that as the absolute values for these parameters are relatively small, the absolute errors of the estimated parameters are acceptable. 

In terms of the goodness-of-fit, the initial log-likelihood (denoted as $\mathcal{LL}_0$) for the null model (with all parameters zero) is $-52,219.17$, the final latent-class model log-likelihood (denoted as $\mathcal{LL}^*$) is $-51,526.13$, and the final baseline model log-likelihood (denoted as $\mathcal{LL}^{\text{B}}$) is $-51,571.67$. We conduct the log-likelihood ratio test \citep{wilks1938large} and obtain the statistic $\chi^2 = -2(\mathcal{LL}^{\text{B}} - \mathcal{LL}^*) = 91.08$, which suggests a p-value of 0 given 5 degrees-of-restrictions (i.e., number of parameters of latent class model minus that of baseline model). This indicates that the latent-class model specification is significantly better than the baseline model. We also calculate the log-likelihood with ``true'' parameters (referred to as $\mathcal{LL}^{\text{True-para}}$) and the value is $-51,535.16$. It is smaller than the $\mathcal{LL}^*$, which means that the estimated parameters have a better goodness-of-fit than the ``true'' parameters. This suggests that the estimation errors may mostly come from random errors in the data generation process, instead of the model estimation.

All parameters have absolute t-values greater than $1.96$, showing significant impacts of these parameters on passengers' path choices. This is reasonable because the synthetic data are generated with those parameters. We also observe that the in-vehicle time shows the highest significance compared to other cost parameters, which is consistent with survey results \citep{jin2017metro}. 

\begin{table}[H]
\caption{Estimation results for the synthetic data}
\centering
\begin{tabular}{@{}ccccc@{}}
\toprule
\multirow{2}{*}{\textbf{Variables}} & \multicolumn{2}{c}{\textbf{Latent class}}               & \multicolumn{2}{c}{\textbf{Baseline}}      \\
                           & Estimation (Error)      & t-value      & Estimation (Error)         & t-value \\ \midrule
\multicolumn{5}{l}{\emph{Choice model}}                                                      \\
In-veh time        & -0.2599 (-2.9\%)       & -10.62        & -0.2254 (-15.8\%)  & -10.03    \\ 
TS: Out-of-veh time            & -0.1991 (-33.2\%)       & -2.88        & -0.3010 (1.0\%)   & -6.88   \\
TS: Num of transfer            & -1.3327 (+1.9\%)        & -4.91        &  -1.8777 (+43.7\%)     & -16.77   \\
TS: Denied waiting time        & -0.3789 (+17.6\%)       & -6.18        & -0.4901 (+52.1\%)   & -18.80   \\ 
CA: Out-of-veh time            &  -0.5419 (-15.14\%)       & -3.82        &-0.3010 (-52.9\%)  & -6.88   \\
CA: Num of transfer            & -2.8071 (-11.5\%)       & -4.98        & -1.8777 (-40.8\%)    & -16.77   \\
CA: Denied waiting time        & -0.7298 (-6.7\%)        & -6.35       & -0.4901 (-37.4\%)   & -18.80   \\ 
$\log PS_m$        & 0.6758 (+16.2\%)      & 12.72        & 0.6054 (+4.1\%) & 12.31  \\ 
$\sigma^k  $         & 0.9191 (-8.1\%)       & 11.05        & 0.9122 (-8.8\%)  & 14.40    \\ 

\midrule
\multicolumn{5}{l}{\emph{Latent group model}}                                                   \\ 

$x^{(1)} $       & 0.6213 (+3.6\%)    & 4.99     & N/A  & N/A   \\ 
$x^{(2)} $        & 1.8637 (+24.3\%)       & 7.01      & N/A   & N/A   \\ 

 \bottomrule
\multicolumn{5}{l}{
\begin{tabular}[c]{@{}l@{}} 
Number of passengers: 2,700. Number of observations: 8,100 \\
$\mathcal{LL}_0$: -52,219.17; $\mathcal{LL}^*$: -51,526.13; $\mathcal{LL}^{\text{B}}$: -51,571.67; $\mathcal{LL}^{\text{True-para}}$: -51,535.16  \\
$\chi^2 = -2(\mathcal{LL}^{\text{B}} - \mathcal{LL}^*) = 91.08$, likelihood ratio test p-value: 0 \\
\end{tabular}} 
\end{tabular}
\label{tab_syn_results}
\end{table}


To further validate the model performance, sensitivity analysis was conducted to explore the impacts of parameter initialization on the model's performance. Moreover, we also evaluate whether the inaccurate estimation of walking speed distribution and left behind distribution would affect the model's performance. 

\begin{figure}[H]
\centering
\includegraphics[width=1 \linewidth]{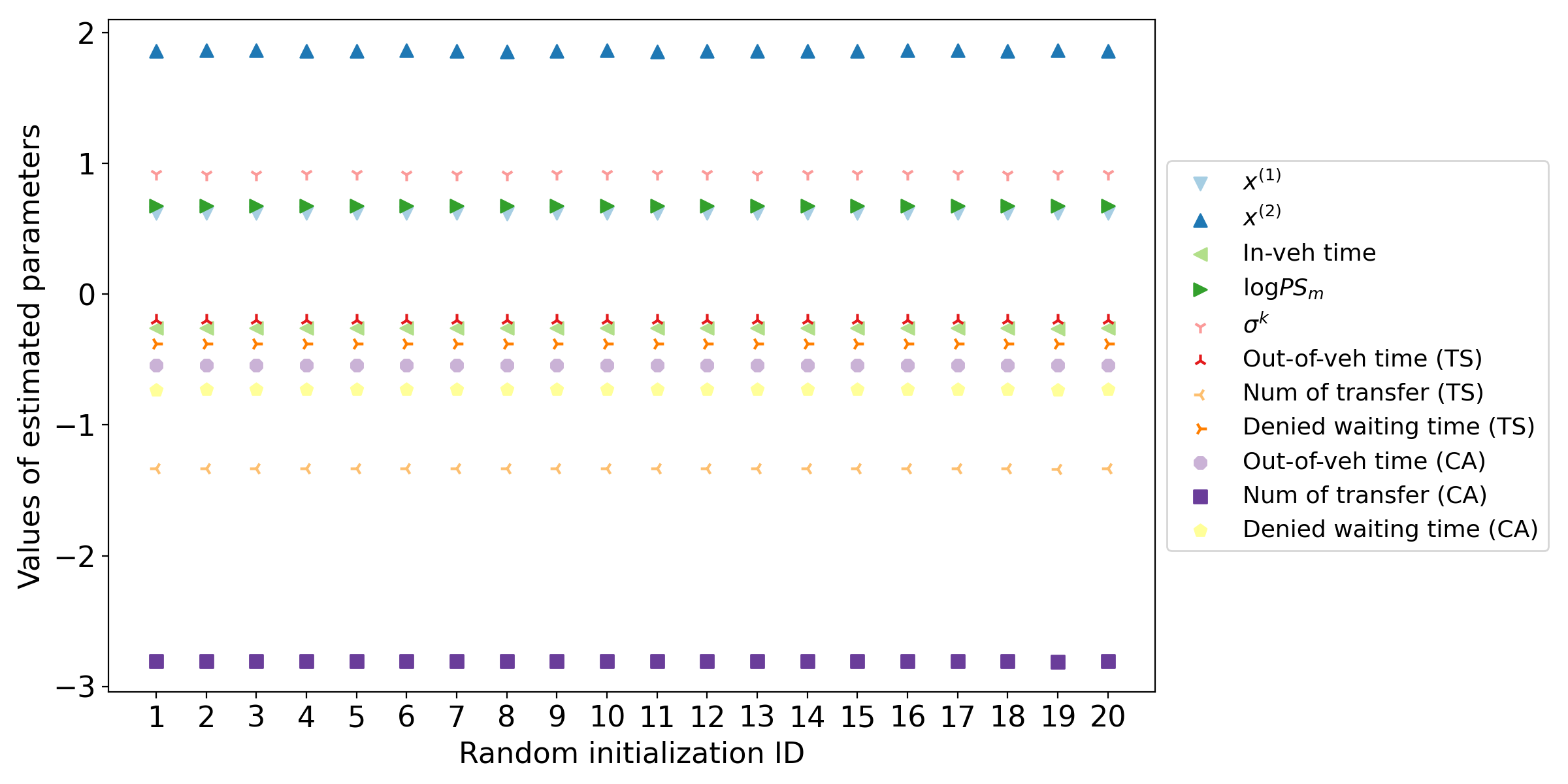}
\caption{Estimated parameters with different initializations}
\label{fig_ini}
\end{figure}

Figure \ref{fig_ini} shows the sensitivity analysis on the initialization of the parameters. A total of 20 experiments are conducted. In each experiment, the initial values of all parameters are drawn uniformly from $\mathcal{U}[-5,5]$. We observe that the final estimated parameters all converged to the same values regardless of initialized parameter values, showing the estimation robustness against the parameter initialization. 

\begin{figure}[H]
\centering
\subfloat[$\mathcal{LL}$ vs. in-veh time and out-of-veh time (TS) ]{\includegraphics[width=0.45\linewidth]{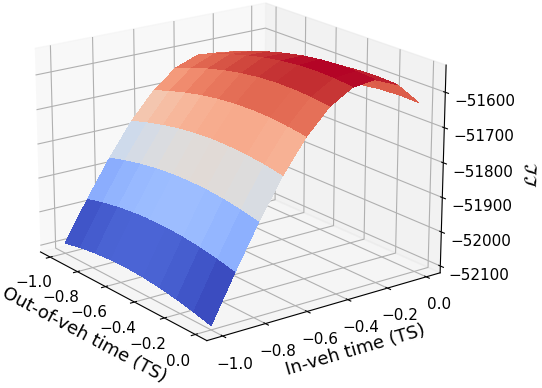}\label{fig_LL1}}
\hfil
\subfloat[$\mathcal{LL}$ vs. $x^{(1)}$ and $x^{(2)}$]{\includegraphics[width=0.45\linewidth]{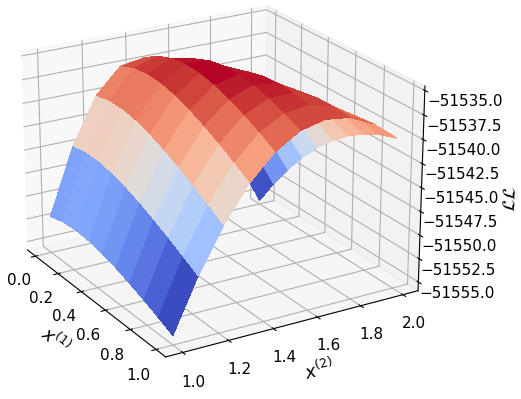}\label{fig_LL2}}
\caption{Log-likelihood function surface}
\label{fig_LL}
\end{figure}

Figure \ref{fig_LL} illustrates the $\mathcal{LL}$ value as a function of variable values. The log-likelihood function is concave around the optimal values\footnote{Due to space limitations, we only show the function curves with respect to four variables}, which further indicates that the estimation results are robust.

\begin{figure}[H]
\centering
\includegraphics[width=0.6 \linewidth]{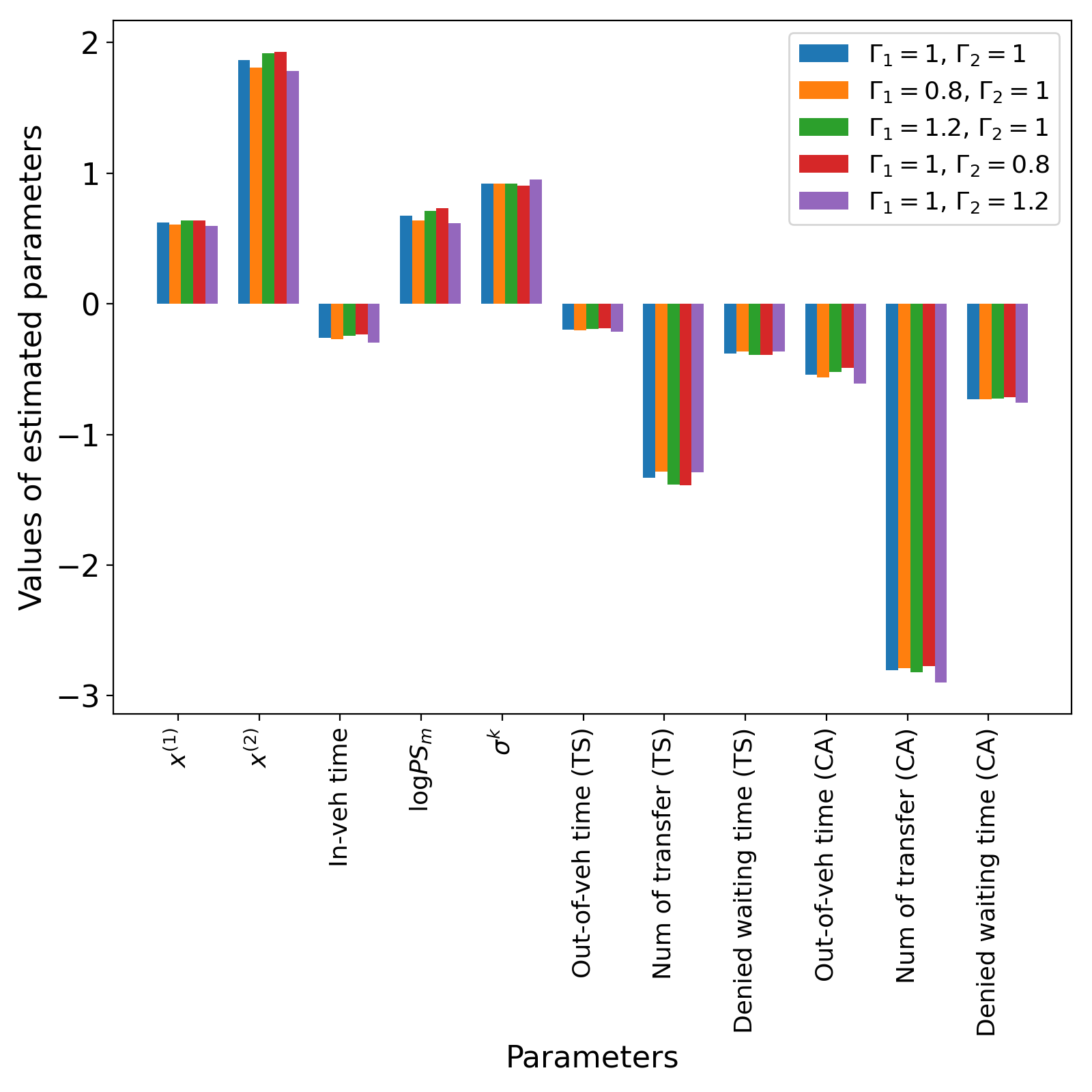}
\caption{Sensitivity analysis on walking speed inputs}
\label{fig_walk_speed}
\end{figure}

Figure \ref{fig_walk_speed} shows the model estimation results with respect to different inputs of walking speeds. We evaluates the model's robustness with respect to errors in walking speed distribution because the estimation passenger's walking speed may not be accurate in the real world. Let $\mu^{\text{WS}}$ and $\sigma^{\text{WS}}$ be the actual walking speed mean and standard deviation when generating the synthetic data (i.e., $\mu^{\text{WS}} = 1.2m/s$ and $\sigma^{\text{WS}} = 0.5m/s$). When estimating the model, we set the speed distribution parameters as $(\Gamma_1 \cdot \mu^{\text{WS}}, \Gamma_2 \cdot \sigma^{\text{WS}})$, where $\Gamma_1, \Gamma_2 \in\{ 0.8, 1, 1.2\}$, which represents different perturbations in the speed parameter inputs ($\Gamma_1=\Gamma_2=1$ means no errors). Figure \ref{fig_walk_speed} shows that the variability of the walking speed distribution does not show much impact on the model performance. 

\begin{figure}[H]
\centering
\includegraphics[width=0.6 \linewidth]{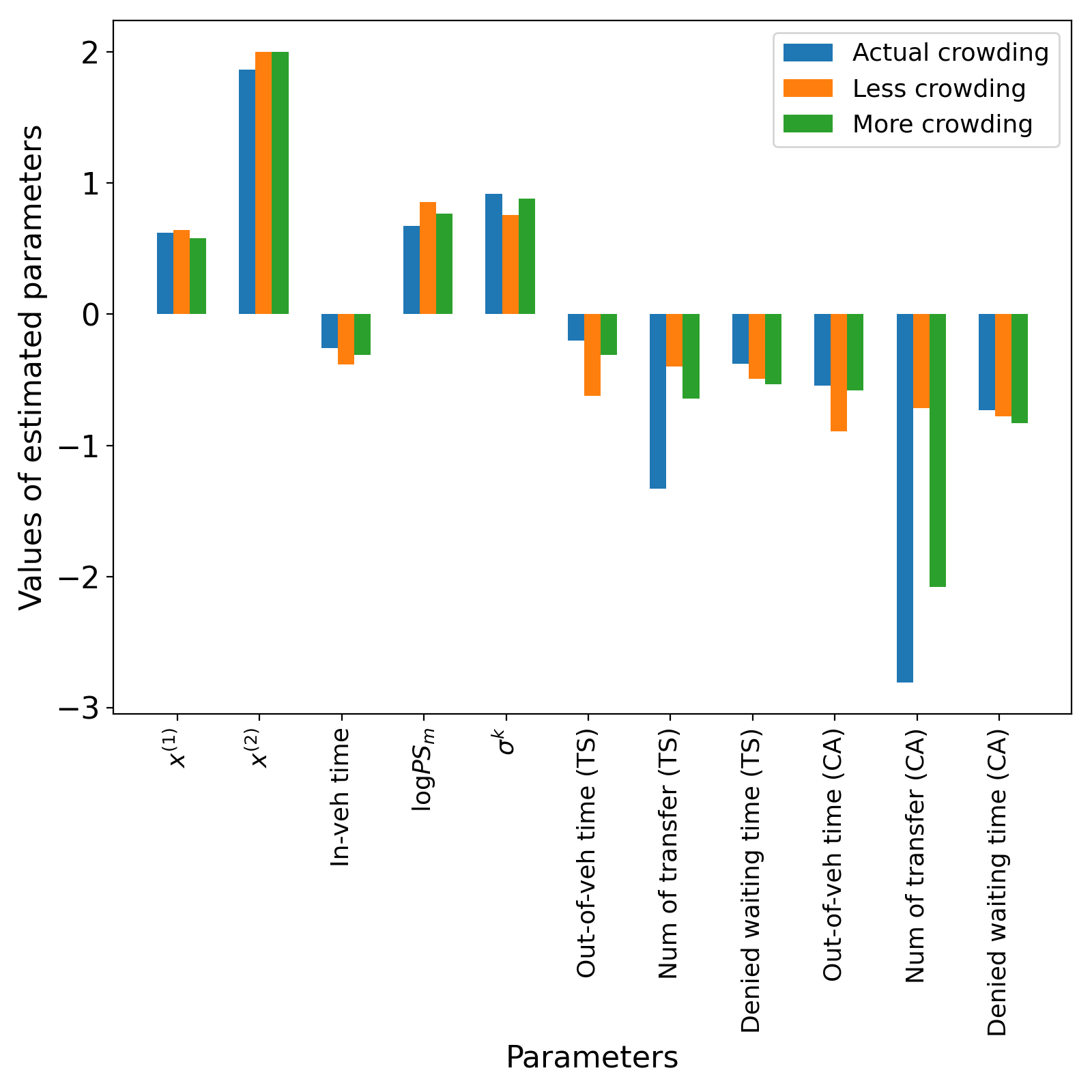}
\caption{Sensitivity analysis on left behind inputs}
\label{fig_db}
\end{figure}

Figure \ref{fig_db} shows the estimation results with respect to different input left behind probabilities at Station C, red line, up direction, which indicates the model's performance if there are estimation errors in left behind probabilities. Similarly, left behind probabilities are chosen for sensitivity analysis because they are values estimated from data and may suffer from errors. Three scenarios are compared: actual crowding (20\% no left behind, 50\% left behind once and 30\% left behind twice, which means no errors), less crowding (80\% no left behind, 20\% left behind once and 0\% left behind twice), and more crowding (10\% no left behind, 20\% left behind once and 70\% left behind twice). It can be seen that the parameters of in-vehicle time, $x^{(1)}$, and  $x^{(2)}$ are not sensitive to the left behind inputs. But other parameters (such as the number of transfers and out-of vehicle time) are highly affected. The reason may be that errors in left behind estimation can affect the model's evaluation of other factors' impacts on travel time.  That is, an additional 10-minute trip time can be caused by more transfers, or high out-of vehicle time, or left behind. If left behind is not estimated accurately, the impact of other factors on total travel time (and passenger choices) may be biased. Hence, the results highlight the importance of incorporating crowding and capacity constraints in the estimation of path choices.

\subsection{MTR empirical case study}
\begin{figure}[htb]
\centering
\includegraphics[width=0.8 \linewidth]{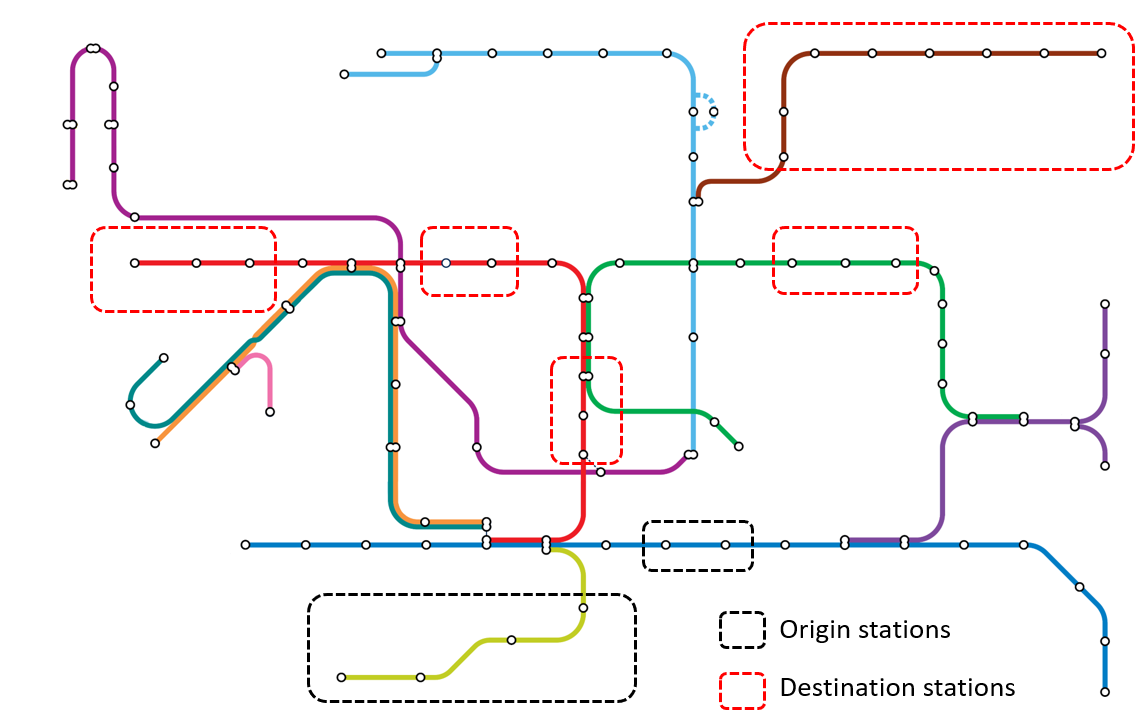}
\caption{Hong Kong MTR network}
\label{fig_hk_net}
\end{figure}

The proposed method is also applied using actual AFC and AVL data from the Hong Kong MTR network. Figure \ref{fig_hk_net} shows the MTR network and select OD pair areas (origins in the black dashed box and destinations in the red). These OD pairs are selected because 1) there are multiple paths between each OD pair, which supports the application of the path choice modeling, and 2) these stations have high enough OD passenger flows to allow the estimation of the left behind probability distribution \citep{ma2019estimation}. 

We randomly select 3,425 passengers with trips between these OD pairs. We consider trips with departure times in the evening peak (5:30 PM - 7:30 PM). Finally, a total of 6,425 trips were collected from the AFC data in July 2018 (i.e., on average each passenger had 1.88 trips). The walking time is assumed to follow the log-normal distribution with mean and variance calibrated by MTR employees. $\alpha^{\text{L}} = -9$, $\alpha^{\text{U}} = 9$, and $\Delta = 1$ are used for numerical integration. 

\begin{table}[H]
\caption{Descriptive statistics of the MTR data}
\centering
\begin{tabular}{lcccc}
\hline
\multicolumn{1}{c}{\textbf{Variables}}               & \textbf{Mean} & \textbf{Std. Dev.} & \textbf{Max} & \textbf{Min} \\ \hline
\multicolumn{5}{l}{\emph{Individual characteristics}}                                                                          \\
\multicolumn{1}{l}{Avg \# travel days per week}      & 4.31          & 2.06               & 7.00         & 0.12         \\
\multicolumn{1}{l}{Std. of 1st trip dept. time (hr)} & 0.68          & 0.51               & 3.54         & 0.01         \\
Min \# stations with 70\% trips                      & 3.22          & 0.15               & 15           & 1            \\
If student (Yes = 1)                                 & 0.10          & 0.30               & 1            & 0            \\\hline
\multicolumn{5}{l}{\emph{Path attributes}}                                                                                     \\
In-veh time (min)                                    & 34.0          & 15.5               & 85.8         & 3.50         \\
Out-of-veh time (min)                                & 9.20          & 1.11               & 24.6         & 0.50         \\
Denied waiting time (min)                            & 1.03          & 2.48               & 16.3         & 0.0          \\ 
 \bottomrule
\multicolumn{5}{l}{
\begin{tabular}[c]{@{}l@{}} 
Number of passengers: 3,425. Number of trips: 6,425 
\end{tabular}} 
\end{tabular}
\label{tab_mtr_data}
\end{table}

We assume passenger's latent groups can be characterized by the following attributes, readily extracted from AFC data: 1) travel frequency, defined as the average number of days with travel per week, 2) schedule flexibility, measured by the standard deviation of the first trip's departure time on weekdays, 3) spatial concentration of trips, defined as the minimum number of stations that covers 70\% of trips in a month, 4) whether the cardholder is a student or not (dummy variable). All these attributes are calculated based on the AFC data in July 2018. The descriptive statistics of the data are shown in Table \ref{tab_mtr_data}. Two latent groups are considered for the experiment. The reason for considering two groups instead of more is that two latent groups are more interpretable in terms of estimation results. The path attributes are the same as the synthetic data experiment except for the ``number of transfers''. The  ``number of transfers'' is dropped from the model due to its high correlation with the ``out-of-vehicle time''. Similar to the synthetic data experiment, we make the parameters of in-veh-time the same for the two groups so that we can compare the scales of out-of-veh time and the number of transfers.

\begin{table}[htb]
\caption{Estimation results for the real-world data}
\centering
\begin{tabular}{ccccc}
\toprule
\multirow{2}{*}{\textbf{Variables}}       & \multicolumn{2}{c}{\textbf{Group 1 (CA)}}            & \multicolumn{2}{c}{\textbf{Group 2 (TS)}} \\
                                 & Estimation           & t-value         & Estimation     & t-value    \\ \hline
\multicolumn{5}{l}{\emph{Choice model}}                                                           \\
In-veh time                  & -0.2785              & -18.13           & \multicolumn{2}{c}{Same as Group 1}       \\

Out-of-veh time                  & -1.1320              & -2.09           & -0.7457        & -9.04      \\
Denied waiting time              & -3.0450              & -1.95           & -0.4517        & -9.90      \\ 
$\log PS_m$              & 1.2611              & 4.38          &  \multicolumn{2}{c}{Same as Group 1}    \\ 
$\sigma^k$             & 2.7294              & 11.99           & \multicolumn{2}{c}{Same as Group 1}     \\ 
\hline

\multicolumn{5}{l}{\emph{Latent group model (Group 2 is set as the base group)}}                     \\ 
ASC$^1$                     & \multicolumn{1}{c}{-0.9644}            & \multicolumn{1}{c}{-6.83}&\multicolumn{2}{c}{0 (fixed)}   \\
Avg \# travel days per week      & \multicolumn{1}{c}{-0.0987}            & \multicolumn{1}{c}{-3.91}&\multicolumn{2}{c}{0 (fixed)}   \\
Std. of 1st trip dept. time      & \multicolumn{1}{c}{0.8667}            & \multicolumn{1}{c}{4.19}&\multicolumn{2}{c}{0 (fixed)}   \\
Min \# stations with 70\% trips & \multicolumn{1}{c}{0.1865}             & \multicolumn{1}{c}{0.40} &\multicolumn{2}{c}{0 (fixed)}   \\
If student (Yes = 1)                      & \multicolumn{1}{c}{1.0192}             & \multicolumn{1}{c}{2.14} &\multicolumn{2}{c}{0 (fixed)}   \\
 \bottomrule
\multicolumn{5}{l}{
\begin{tabular}[c]{@{}l@{}} 
$^1$: ASC: alternative specific constant \\ 
Number of passengers: 3,425. Number of observations: 6,425 \\
$\mathcal{LL}_0$: -32,157.84; Final $\mathcal{LL}^*$: -30,867.48 \\
$\chi^2 = -2(\mathcal{LL}_0 - \mathcal{LL}^*) = 2,580.72$, likelihood ratio test p-value: 0\\
\end{tabular}} 
\end{tabular}
\label{tab_real_results}
\end{table}

The estimation results are shown in Table \ref{tab_real_results}. The two latent groups are referred to as Group 1 and 2, respectively. Group 2 is set as the base alternative in the group-assignment multinomial logit model (Eq. \ref{eq_group}). Results show that the signs and scales of all parameters are reasonable. All time-related parameters are significantly negative. The value of the in-vehicle time parameter is -0.2785, which is similar to the survey results (-0.2676) \citep{jin2017metro}. For both Groups 1 and 2, the absolute values of the parameters for out-of-vehicle time and denied waiting time are both greater than that of the in-vehicle travel time, reflecting that passengers were more sensitive to walking and waiting times compared to the time seated in vehicles. And these results are also consistent with the survey. 

Comparing the results of Group 1 and 2, we observe that the out-of-vehicle and denied waiting times show a larger impact on the path choice utility for passengers in group 1, which implies that Group 1 is possibly comfort-aware (CA) and Group 2 is time-sensitive (TS). The parameters determining the latent groups indicate that CA passengers have less travel frequency (i.e., the effect of avg. \# travel days per week is negative) and more schedule flexibility (i.e., the effect for the std. of the 1st trip departure time is positive), and students are more likely to belong to this group. These suggest that CA passengers are most likely to be irregular users and they may mainly use the MTR system for non-work activities (such as entertainment). Hence, they care more about the trip comfort and prefer paths with less walking and waiting times. In contrast, TS passengers have higher travel frequency and less schedule flexibility and they may use the metro system mostly for regular commuting trips. Besides, they care more about saving the total travel time to arrive at their destinations on time. The spatial concentration variable (i.e., min \# stations with 70\% trips), though having a positive impact on being in the CA group, is not statistically significant (t-value 0.4). $\sigma^k$ is significant for both groups, which means that the panel effects are diverse across populations (i.e., some have more stable travel patterns but are not).

\section{Conclusion}
Understanding passenger path choices are important for operations management in urban rail systems, especially those with crowded conditions. This paper presents a probabilistic approach for the path choice model estimation with train capacity constraints using AFC (tap-in and tap-out) and AVL data. The choice heterogeneity and longitudinal behavioral correlations are captured by a latent class model with panel effects. Passenger's movement is formulated using a passenger to train assignment model with explicit modeling of the processes of access/egress, left behind (crowding), and transfer. A tractable likelihood function is derived to facilitate the model estimation. The t-value of estimated parameters is calculated based on the numerically estimated Hessian matrix and Cramer–Rao bound. The method is data-driven, flexible to accommodate different choice models, and easy to solve using non-linear optimizers. 

The model performance is validated using synthetic data to estimate the individual choice parameters. The sensitivity analysis affirms its robustness to parameter initialization and small errors in inputs (walking speed and left behind distributions). It also highlights that neglecting capacity constraints (left behind) can lead to biased estimation of path choice parameters under crowding conditions. The model is also applied using real-world data from the MTR system in Hong Kong. It reveals two different groups of passengers: time-sensitive (TS) and comfort-aware (CA). TS passengers are generally regular commuters with high travel frequency and small schedule flexibility. They are more likely to choose paths with less trip travel times. CA passengers care more about travel comfort and prefer choosing paths with less walking and waiting times. 

Example policy implications can be derived from the case study. As these two groups of passengers value in-vehicle and out-of-vehicle times differently, transit agencies can use this insight to design customized route recommendation systems with better passenger acceptance. Moreover, route recommendations can help to relieve congestion by recommending TS and CA passengers to use different routes during peak hours. Interesting future research directions include: 1) Explore the evolution of choice preferences and learning behavior over time under network interventions (such as network extension and demand management policies). 2) Develop a downstream model to utilize the latent passenger group information for better route recommendations or fare policies.

\section{Authors’ contribution}
\textbf{Baichuan Mo}: Conceptualization, Methodology, Software, Formal analysis, Data Curation, Writing - Original Draft, Writing - Review \& Editing, Visualization.  \textbf{Zhenliang Ma}: Conceptualization, Methodology, Supervision, Formal analysis, Data Curation, Writing - Original Draft, Writing - Review \& Editing. \textbf{Haris N. Koutsopoulos}: Conceptualization, Supervision, Formal analysis.  \textbf{Jinhua Zhao:} Conceptualization, Supervision, Project administration, Funding acquisition.

\section{Acknowledgement}
The authors would like to thank Chicago Transit Authority (CTA) for their support and data availability for this research. 

\appendix
\appendixpage
\section{Left behind probability calibration}\label{sec_LB_est}
\begin{figure}[htb]
\centering
\includegraphics[width=0.5 \linewidth]{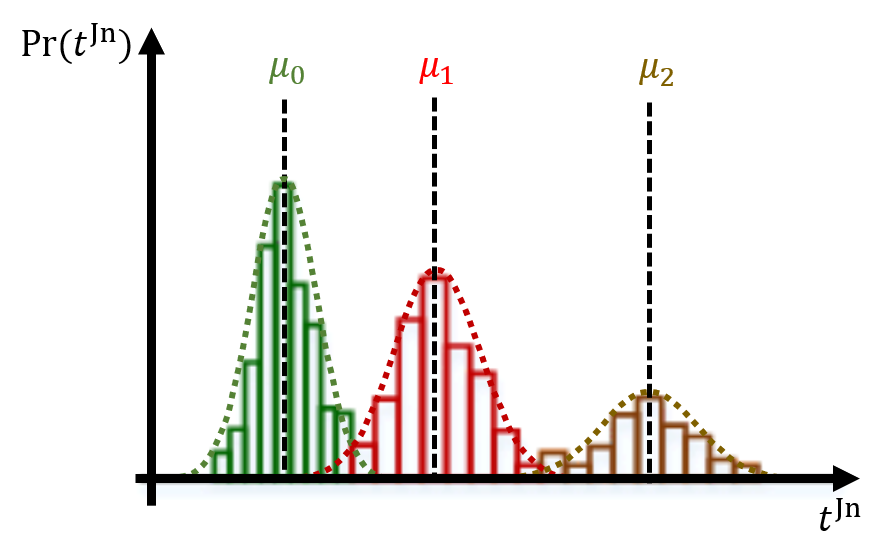}
\caption{Example of journey time distribution}
\label{fig_journey_time}
\end{figure}

The left-behind probability can be estimated from a Gaussian mixture model proposed by \citet{ma2019estimation}. The main idea is that passengers being left behind different times would have different journey times. Let $t^{\text{Jn}}_i$ be the random variable indicating the journey time of passenger $i$ (i.e., tap-out time minus the tap-in time). Figure \ref{fig_journey_time} shows an example of journey time distribution between a specific OD pair. We observe there are three clusters, indicating passengers being left behind 0, 1, and 2 times at the origin station. Hence, we can model the journey time distribution as a Gaussian mixture model:
\begin{align}
\text{Pr}(t^{\text{Jn}}_i; \boldsymbol{\mu},\boldsymbol{\sigma},\boldsymbol{w}) = \sum_{c=0}^{C} w_c \cdot \Phi(\mu_c, \sigma_c)
\end{align}
where $w_c$ is the (unknown) fraction of passengers in cluster $c$, $w_c > 0$ and $\sum_{c=0}^{C} w_c = 1$. $C$ is the total number of clusters (i.e., the maximum number of left behind times at the origin station). $\Phi(\mu_c, \sigma_c)$ is the PDF for the Gaussian distribution $\mathcal{N}(\mu_c, \sigma_c)$.

The Gaussian mixture model can be estimated by solving the following problem:
\begin{subequations}
\begin{align}
  \max_{\boldsymbol{\mu},\boldsymbol{\sigma},\boldsymbol{w}} \quad & \sum_{i\in \mathcal{P}'} \log (\text{Pr}(t^{\text{Jn}}_i;\boldsymbol{\mu},\boldsymbol{\sigma},\boldsymbol{w}))   \\
    \text{s.t.} \quad
    & \text{Auxiliary Constraints} \\
    & \sum_{c\in C} w_{c} = 1  \\
    & w_{c} \geq 0 \quad \forall c = 0,1,...,C
\end{align}
\label{eq_lb}
\end{subequations}
where  $\mathcal{P}'$ is the passenger set used for the left behind probability estimation. The auxiliary constraints are used for model stability. These constraints contain prior human knowledge on the journey time distribution, such as the difference between $\mu_c$ and $\mu_{c + 1}$ should be close to a headway, the mean journey time without being left behind (i.e., $\mu_0$) should be close to the sum of access, egress, and in-vehicle times.  More information on the model can be found in \citet{ma2019estimation}.

The model is station and time-specific, which enables the calibration of left behind probabilities for each station at different time intervals in the system (by adjusting $\mathcal{P}'$). The estimated $w_c$ (i.e., the fraction of passengers in cluster $c$) is the probability of being left behind $c$ times at the corresponding station and time period. 

\section{Numerical calculation of Hessian matrix}\label{sec_hessian}

According to \citet{fornberg1988generation}, given a general function $f(x,y)$, the second derivative with a fourth-order accuracy can be calculated as
\begin{align}
& \frac{\partial^2 f(x,y)}{\partial x^2}|_{x_0,y_0} = \frac{1}{h_x^2}\left[ \frac{-1}{12}f(x_{-2},y_0) + \frac{4}{3}f(x_{-1},y_0)  \right. \nonumber \\
&  \left. + \frac{-5}{2}f(x_{0},y_0) + \frac{4}{3}f(x_{+1},y_0) + \frac{-1}{12}f(x_{+2},y_0)\right] + \mathcal{O}(h_x^4) 
\label{eq_hess1}
\end{align}
and
\begin{align}
& \frac{\partial^2 f(x,y)}{\partial x \partial y}|_{x_0,y_0} = \frac{1}{h_x h_y}\left[ \frac{-1}{48}f(x_{-2},y_{-2}) + \frac{1}{3}f(x_{-1},y_{-1})  \right. \nonumber \\
&   + \frac{-1}{3}f(x_{-1},y_{+1})+ \frac{-1}{3}f(x_{+1},y_{-1}) + \frac{1}{3}f(x_{+1},y_{+1})  \nonumber \\
& \left.+ \frac{1}{48}f(x_{+2},y_{-2}) + \frac{1}{48}f(x_{-2},y_{+2})  + \frac{-1}{48}f(x_{+2},y_{+2})\right] \nonumber \\
& + \mathcal{O}(h_x^2h_y^2)
\label{eq_hess2}
\end{align}
where $h_x$ and $h_y$ are small perturbations for $x$ and $y$, respectively. $x_k$ ($y_k$) represents $x_0+kh_x$ ($y_0+kh_y$). The derivation is based on Taylor's series expansion with uniform grid spacing \citep{fornberg1988generation}.

\bibliography{mybibfile}

\end{document}